\documentclass[12pt,a4paper]{article}
\usepackage{a4wide}
\usepackage[centertags]{amsmath}
\usepackage{amssymb}
\usepackage{color}
\usepackage{epsfig}
\usepackage{varioref}
\usepackage{cite}
\usepackage[tight]{subfigure}
\usepackage{ulem} 

\newcommand{\bs}[1]{\boldsymbol{#1}}
\newcommand{\SU}[1]{\text{SU}(#1)}
\newcommand{\U}[1]{\text{U}(#1)}
\newcommand{\m}{\text{-}}
\newcommand{\susy}{{\textsc{susy}}}
\newcommand{\alphagut}{\ensuremath{\alpha_{\textsc{gut}}}}
\newcommand{\mgut}{\ensuremath{M_{\textsc{gut}}}}
\newcommand{\sgn}[1]{\ensuremath{\text{sgn\,}\mu}}
\newcommand{\Root}{\texttt{ROOT}}
\newcommand{\softsusy}{\texttt{SOFTSUSY}}
\newcommand{\indisoft}{\texttt{INDISOFT}}
\newcommand{\cpp}{\texttt{C++}}
\newcommand{\trace}[1]{\ensuremath{{\rm Tr}\left\{#1\right\}}}
\newcommand{\code}[1]{\small{\texttt{#1}}\normalsize}

\labelformat{equation}{Eq.~(#1)} 
\labelformat{figure}{Fig.~#1} 
\labelformat{subfigure}{Fig.~#1} 
\labelformat{table}{Tab.~#1} 

\begin{document}

\setlength{\parindent}{0mm}

\thispagestyle{empty}

\begin{center}
{\huge Four Generations: \susy{} and \susy{} Breaking}
\vspace*{5mm} \vspace*{1cm}
\end{center}
\vspace*{5mm} \noindent
\vskip 0.5cm
\centerline{\bf Rohini M.~Godbole, Sudhir K.~Vempati, Ak\i{}n Wingerter}
\vskip 1cm
\centerline{
\em Centre for High Energy Physics,}
\centerline{\em Indian Institute of Science, Bangalore 560012, India}
\vskip2cm

\centerline{\bf Abstract}
\vskip .3cm

We revisit four generations within the context of supersymmetry. We
compute the perturbativity limits for the fourth generation Yukawa
couplings and show that if the masses of the fourth generation lie
within reasonable limits of their present experimental lower bounds,
it is possible to have perturbativity only up to scales around 1000
TeV. Such low scales are ideally suited to incorporate gauge mediated
supersymmetry breaking, where the mediation scale can be as low as
10-20 TeV. The minimal messenger model, however, is highly
constrained. While lack of electroweak symmetry breaking rules out a large part of the parameter space,
a small region exists, where the fourth generation stau is
tachyonic. General gauge mediation with its broader set of boundary
conditions is better suited to accommodate the fourth generation.

\vskip .3cm

\clearpage
\newpage
\section{Introduction}

Interest in a sequential fourth generation of chiral fermions has waxed and 
waned over the past decades with the changing status of constraints implied by various precision measurements in the flavour and the gauge sector.  The recent observation of single-top events at the Tevatron~\cite{Abazov:2009ii,Aaltonen:2009jj} allowed a direct and clean determination of the CKM matrix element $|V_{tb}| = 0.91 \pm 0.11 \,\text{(stat+syst)} \pm 0.07 \,\text{(theory)}$~\cite{Aaltonen:2009jj}. This is in good agreement with the Standard Model prediction of $|V_{tb}| \simeq 1$, but falls short of excluding extensions of the theory by another chiral generation of quarks and leptons, as has been emphasised in recent publications \cite{Alwall:2006bx,Kribs:2007nz,Bobrowski:2009ng,Chanowitz:2009mz,Holdom:2009rf,Novikov:2009kc}. As a matter of fact, the mixing between the third and a hypothetical fourth family\footnote{We will denote the fourth generation quarks and leptons by $t'$, $b'$, $\tau'$, $\nu_\tau'$. For enhanced readability, in graphs we may use the alternate notation $t4$, $b4$, $\tau4$, $\nu_{\tau4}$.} can be as large as the mixing between the first two generations in the Standard Model and yet be compatible with all experimental data including direct searches for new quarks and leptons, electroweak precision measurements, and flavour changing neutral currents \cite{Hung:2007ak,Bobrowski:2009ng,Chanowitz:2009mz,Kribs:2007nz,Novikov:2009kc}.

\medskip

Currently, we do not have a theoretical understanding of the number of 
generations, and a priori there is no reason why there should not be another 
one. A fourth generation would have profound implications for particle physics 
phenomenology\footnote{For a review, see Ref.~\cite{Frampton:1999xi}.}. For one
thing, it can ease the tension between the LEP bound on the Higgs mass 
and the electroweak precision 
measurements~\cite{Kribs:2007nz,Chanowitz:2009mz,Novikov:2009kc} as we will elaborate on in Section \ref{sec:limits}. 
Secondly, the Yukawa couplings of the fourth generation fermions, 
heavier than those of the first three, will have important implications 
for the perturbativity of the theory at the high scale. This can have 
nontrivial effects, for example, on the Higgs mass bounds obtained in the 
SM by demanding that the Landau pole lie above the Planck 
scale~\cite{Nielsen:1995gx,Pirogov:1998tj}. As a result of this possible 
effect of the fourth generation on the consistency of the theory up to  
high scale, unified theories with four generations have also
received special attention, both with supersymmetry (SUSY)~\cite{Goldberg:1985gj,Enqvist:1985ct,Arnowitt:1987xk,Drees:1987ev,Gunion:1994zm,Gunion:1995tp,Dubicki:2003am,Murdock:2008rx} 
and without it~\cite{Hung:1997zj}.  The fourth generation can also account 
for the extra CP violation needed for electroweak baryogenesis to 
work \cite{Hou:2008xd,Fok:2008yg,Kikukawa:2009mu}. Other ideas that have 
been explored in connection with four generations include: Extra 
dimensions~\cite{DePree:2009ed,Burdman:2007sx,BorstnikBracic:2006xc}, 
technicolour~\cite{Stremnitzer:1987zp,Frandsen:2009fs,Antipin:2009ks} 
and electroweak symmetry breaking~\cite{Holdom:1986rn,Hill:1990ge,King:1990he}, radiative mass generation~\cite{Kramer:1981sq,Kagan:1989fp}, bounds from 
cosmology~\cite{Sher:1992yr}, 
neutrino physics~\cite{Fritzsch:1992bv,Hill:1989vn,Babu:2009aq}, and 
finally string theory~\cite{Drees:1986ug}.

\medskip

Recent analyses of the fourth generation have mainly focused on the the non-supersymmetric case. In this, we extend the minimal supersymmetric standard model (MSSM) by one chiral generation and explore the implications for supersymmetry and supersymmetry breaking. For clarity, we will call the MSSM with three and four generations MSSM3 and MSSM4, respectively. In Section \ref{sec:limits}, we review the constraints on the masses of the fourth generation coming from experiment, precision electroweak data, and flavour changing neutral currents. In Section \ref{sectwo}, we introduce our notation and map out the parameter space of the MSSM4 where the theory remains perturbative up to some assumed unification scale of the order $2\times10^{16}$ GeV. We will find that this puts severe restrictions on $m_{t'}$, $m_{b'}$, $m_{\tau'}$, $\tan\beta$, and show that the current experimental bounds and perturbative unification are mutually exclusive. Depending on the masses and $\tan\beta$, the theory becomes strongly-coupled around 10-1000 TeV. To illustrate the qualitative differences, we will present a toy mSUGRA model\footnote{Here and in the following, we will use the word mSUGRA, where we should more correctly call it the constrained minimal supersymmetric standard model (CMSSM).} where we have chosen the quark and lepton masses to be equal to their third generation counterparts. In Section \ref{sec:gmsb}, motivated  by the low ``perturbativity'' scale, we explore this issue in the context of gauge mediated supersymmetry breaking (GMSB). Minimal GMSB, however, suffers from tachyons in the spectrum, and hence we are led to generalise our GMSB set-up in the quest for realistic models. Finally, in Section \ref{sec:summary} we summarise our results and outline directions for future work.


\section{Limits from Experiment, Electroweak Precision Data, and FCNC}
\label{sec:limits}

The current experimental limits quoted by the PDG \cite{Amsler:2008zzb} at 95\% CL~are :
\begin{equation}
m_{t'} \gtrsim 256 \text{ GeV,} \quad m_{b'} \gtrsim 128 \text{ GeV,} \quad m_{\tau'} \gtrsim 100.8 \text{ GeV,} \quad m_{\nu_{\tau}'} \gtrsim 45 \text{ GeV.}
\label{eq:experimental_limits_on_masses}
\end{equation}
The bounds on $m_{t'}$ and $m_{b'}$ assume that the predominant decay mode is to a $W$ boson and another quark \cite{:2008nf,Abachi:1995ms}, which we expect to be true for a sequential fourth generation\footnote{One can find a higher lower bound on $m_{b'}$ by adding model-specific assumptions \cite{Aaltonen:2007je,Acosta:2002ju}.}. Along the same lines, for the limit on $m_{\tau'}$ one assumes that $\tau'$ decays to a $W$ and a stable $\nu_{\tau}'$ \cite{Achard:2001qw}, and the limits for a stable heavy neutral particle give the lower bound on $m_{\nu_{\tau}'}$ \cite{Abreu:1991pr}.

\medskip

Electroweak precision measurements further constrain the allowed mass ranges. For no mixing between the third and fourth generation, $\chi^2$ is minimised for $|m_{t'} - m_{b'}| \simeq 45 - 75$ GeV \cite{Kribs:2007nz}, and for $m_{t'} = 300$ GeV with  $m_{b'}$ subject to this constraint, the mixing can be as large as $\sin\theta_{34} = 0.35$ \cite{Chanowitz:2009mz}. The precision data excludes larger mixing between the third and fourth families \cite{Chanowitz:2009mz} that is otherwise allowed from FCNC constraints \cite{Bobrowski:2009ng}.

\medskip

Fig.~10.4 of Ref.~\cite{Amsler:2008zzb} by the PDG shows the constraints from electroweak precision measurements. The values $S = -0.04 \pm 0.09$ and $T = 0.02 \pm 0.09$ obtained from the fit are in best agreement with a Higgs mass of $m_H=117$ GeV. For higher Higgs masses, the 90\% CL contours move towards smaller $S$ and larger $T$ values. From this it is clear that a fourth generation may ease the tension between the LEP bound on the Higgs mass and the electroweak precision data, since its contribution to $T$ is typically positive \cite{Kribs:2007nz,Chanowitz:2009mz}. A fourth generation \textit{can} give a negative contribution to $S$, but the mass splitting of the quarks is constrained by the $T$ parameter and leads to a positive correction to $S$ that has to be kept small in order to stay in the 90\% CL ellipse.

\medskip

Note that the limits quoted in \vref{eq:experimental_limits_on_masses} are at 95$\%$ CL.
In the absence of direct availability of these bounds at a higher level of
confidence, we approximate them  by subtracting 20\% off the respective limits
at 95$\%$ CL. Furthermore, the exclusion limits denote the pole masses, whereas
in our calculations, we need to use the running masses. We will account for 
this difference by taking yet another 5\% off the masses for QCD corrections. Note that
this is a conservative estimate as the SUSY threshold corrections can induce an additional
difference up to 20\% between the pole and the running masses \cite{Rattazzi:1995gk}.

\medskip

In our analysis, we will thus be working with two sets of 
mass limits, namely those at $95\%$ CL as well as the weaker lower bounds obtained with
the above-mentioned prescription. Both sets are subject to the constraint from 
electroweak precision measurements that the mass splitting in the 
same \SU{2} multiplet be  not greater than $\sim$ 75 GeV. Thus we consider in our analysis the following sets of fourth generation fermion masses:
\begin{equation}
m_{t'} = 256 \text{ GeV,} \quad m_{b'} = 181 \text{ GeV,} \quad m_{\tau'} = 100.8 \text{ GeV}
\label{pdglimits}
\end{equation}
\begin{equation}
m_{t'} = 192 \text{ GeV,} \quad m_{b'} = 117 \text{ GeV,} \quad m_{\tau'} = \phantom{1.8}75 \text{ GeV}
\label{ourlimits}
\end{equation}
We will comment on the mass of the fourth generation neutrino in Section \ref{sectwo}.


\section{Perturbativity  and Four Generations}
\label{sectwo}

One of the main constraints in considering models with four generations  
is the perturbativity of the Yukawa couplings.  With the masses of the 
fourth generation expected to be typically larger than the already known 
third generation masses, the scale up to which the theory remains 
perturbative is a major concern.  It is expected that supersymmetry would 
soften the running of the Yukawa couplings and enable the theory to be valid 
up to much higher energy scales. It is well known that 
the Yukawa couplings of the heavy fermions flow under the renormalisation group to a infrared (quasi-)fixed point. 
The proximity of the observed top mass to this limit
(see e.g.~Ref.~\cite{Ananthanarayan:1991xp}), in fact also  indicates that 
a fourth generation of fermions, heavier than the top quark, can have 
important implications for the high scale perturbativity of the theory.

\medskip

In the present section, 
we discuss this issue in detail and present the results of our computations. 
As we will show, these high scales will extend only up to $\sim$ 1000 TeV. 
 Let us also note that in supersymmetric theories the scale up to which 
perturbativity is preserved has implications for (a) gauge coupling unification
and (b)  the scale of supersymmetry breaking, though \textit{a priori} they 
are two independent sectors of the theory. 

\medskip

The generalisation of the MSSM to four generations is straight-forward. We will denote the fourth
generation as the primed generation: ($t',b',\tau',\nu_\tau'$) and denote this model as MSSM4. The standard MSSM picture
with three generations will from now on be mentioned as MSSM3.  The MSSM4 superpotential takes the form :
\begin{equation}
\label{superpotential}
W = u^c ~ \mathbf{Y_u} Q H_2 + d^c ~\mathbf{Y_d} Q H_1 + e^c~ \mathbf{Y_e} L H_1 + \mu H_1 H_2 ,
\end{equation}
where  $\mathbf{Y}$ are $4 \times 4$ matrices in generation space. The Yukawa couplings  of
the fourth generation, defined in terms of their masses,  are given as 
\begin{equation}
\label{yukawas}
h_{t'} = ~{m_{t'}  \sqrt{2}  \over v \sin \beta   } \;\; ,\;\; h_{b'} = ~{m_{b'} \sqrt{2} \over    v \cos \beta } \;\;,\;\;  
h_{\tau'} =~{m_{\tau'}  \sqrt{2} \over  v \cos \beta},
\end{equation}
where $v=246$ GeV stands for the Higgs vacuum expectation value.

\medskip

In addition, the fourth generation neutrino $\nu_{\tau'}$ should also attain a 
mass $m_{\nu_\tau'} ~> 45$  GeV.  There are several ways of generating 
the fourth generation neutrino mass term , which may either be of the Dirac
type or a mixture of Dirac plus Majorana type, leading to a Majorana mass for 
the $\nu_{\tau'}$ and this can make the analysis highly model dependent. In 
the present case, we have not considered the effects of a Dirac Yukawa coupling
for the neutrino in the renormalisation group equations (RGE). It should be noted that a large 
neutrino Dirac Yukawa coupling can, in principle,  affect strongly the 
evolution of the  Yukawa coupling of the $\tau'$  but, the effect would be 
minimal as long as the said neutrino Dirac Yukawa coupling is small.  
A more detailed analysis of the various possible neutrino Yukawa 
couplings and their impact on perturbativity will be presented elsewhere. 

\medskip

Firstly,  let us  note that the larger masses of the down type fourth
generation fermions mean that requiring that the Yukawa couplings be 
perturbative at the weak scale puts upper bounds on tan~$\beta$, stronger
than the ones present in the MSSM3. The strongest limit\footnote{Lower limits
on $m_{t'}$ on the other hand put lower limits on $\tan\beta$ of limited 
consequence.} comes from $h_{b'}$. 
Imposing that $h_{b'}^2 \sim 4\pi$, we have :
\begin{equation}
\tan\beta \leq \left(2 \pi \left({v \big/ m_{b'}} \right)^2 -1 \right)^{1\over 2}
\end{equation}
For the lower limit of $m_{b'}$ quoted in \vref{pdglimits} this already sets a limit  $\tan\beta~ \leq$ 4.7.
 The bound from $m_{\tau'}$ is much weaker, $\tan\beta \sim $ 38.  
Even for the values of $\tan\beta$ obeying this limit, the Yukawa couplings are 
not expected to be perturbative all the way up to the GUT scale, 
where the gauge coupling unification happens. 

\medskip

It should be noted that in the presence of four generations, gauge coupling 
unification at the 1-loop level still takes place at the scale 
$M_{\text{GUT}} \sim 2 \times 10^{16}$ GeV, though the value of 
$\alphagut$ itself changes.  We refer to 
Appendix~\ref{appendix-gaugecouplings} for the relevant renormalisation group 
equations.  At the 2-loop level, the renormalisation group running of 
gauge couplings will involve Yukawa couplings and thus if the Yukawa 
couplings become non-perturbative at scales much lower than $\mgut{}$ 
they could render the same to the gauge couplings. In the non-supersymmetric 
Standard Model with four generations, it has been known that perturbative 
unification of the gauge couplings is possible \cite{Hung:1997zj}.  However, 
the values for the fourth generation masses used in the  analysis of Ref.~\cite{Hung:1997zj} are ruled by the recent experimental results (see also Ref.~\cite{Pirogov:1998tj}). 
 
\medskip

In the following, we will derive upper limits on the fourth generation fermion masses by requiring that the 
Yukawa couplings of the  fourth generation be perturbative all the way up to the GUT scale.\footnote{Related analysis can be found in Refs.~\cite{Dubicki:2003am,Murdock:2008rx}.} 

\medskip

For this purpose, we use the two loop RGE equations listed in Appendix \ref{appendix-yukawarge} for $n_G$
number of generations. Needless to say, $n_G =3 $ for MSSM3 and $n_G =4$ for MSSM4.  We solve these
equations numerically for a given set of masses at the weak scale and check whether the corresponding 
Yukawa couplings remain perturbative at the high scale.

\medskip

We have developed a variant of the popular supersymmetric spectrum calculator,
\softsusy{} \cite{Allanach:2001kg}, called \indisoft{}. It contains significant modifications including
the ability to  handle three or four generations. Some technical aspects are presented in the
Appendix \ref{app:indisoft}.

\medskip

We first vary $m_{t'}$ and $m_{b'}$ while keeping $m_{\tau'}$ fixed and later vary $m_{b'}$ and $m_{\tau'}$ while keeping $m_{t'}$ fixed.   
From the RGE in Appendix \ref{appendix-yukawarge} we see that the evolution of $t'$ 
is independent of the $\tau'$ mass at 1-loop order.  With this in mind, for the
first analysis we take the $\tau'$ mass to be negligible. For numerical purposes
we set it equal to the $\tau$ mass, i.e.~$m_{\tau'} ~=m_{\tau}$.  Before presenting the numerical results one final comment is in 
 order. The Yukawa couplings of the heavy fields (e.g.~$t'$, $t$) flow towards an infrared fixed point at the 
 weak scale.  Using this one can derive an upper bound on these masses of the heavy fermions analytically.  
The analytical results for the $t'$ are presented in Appendix \ref{appendix-upperbounds}.

\begin{figure}[p]
\centering
\subfigure{\includegraphics[width=0.45\textwidth]{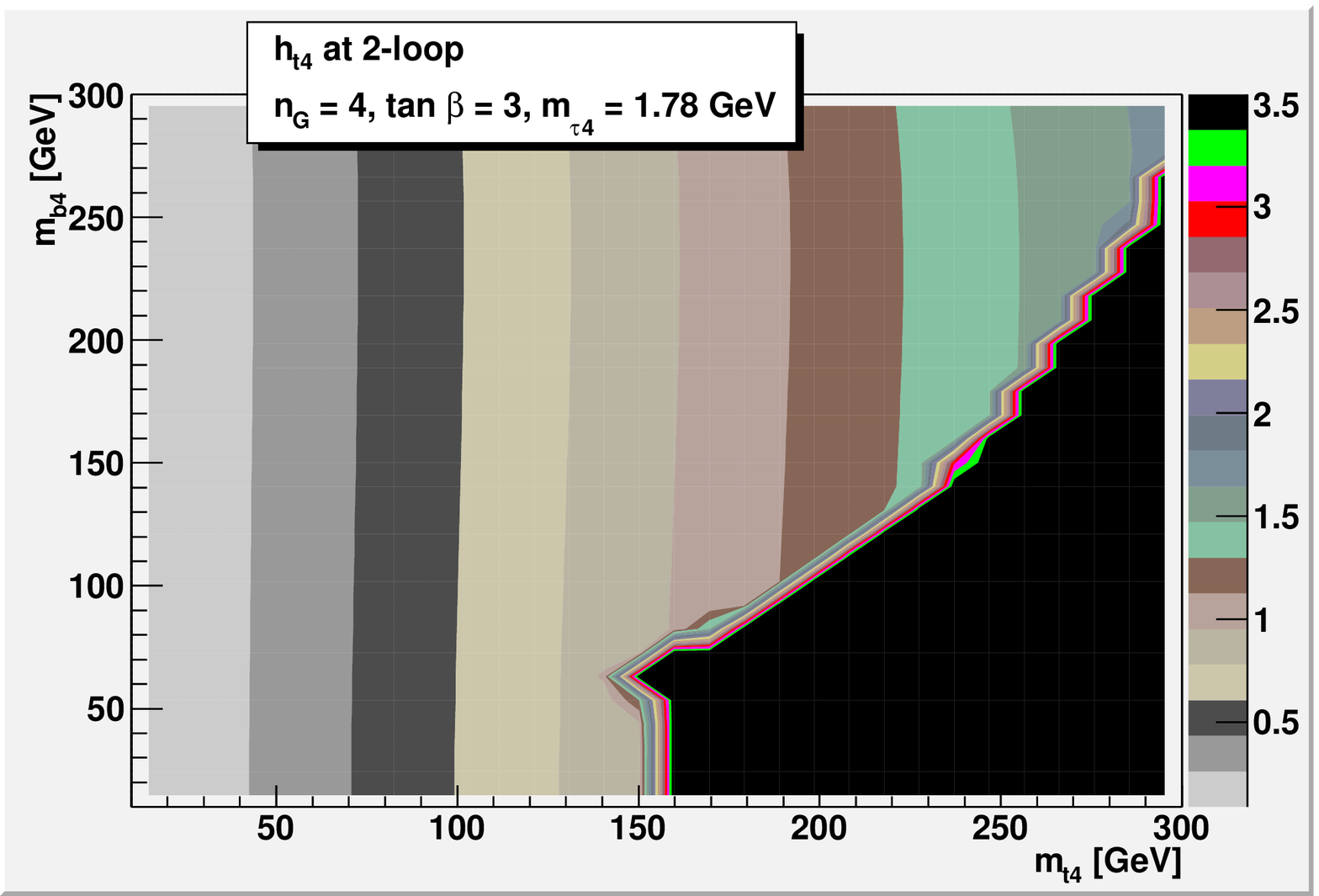}}
\subfigure{\includegraphics[width=0.45\textwidth]{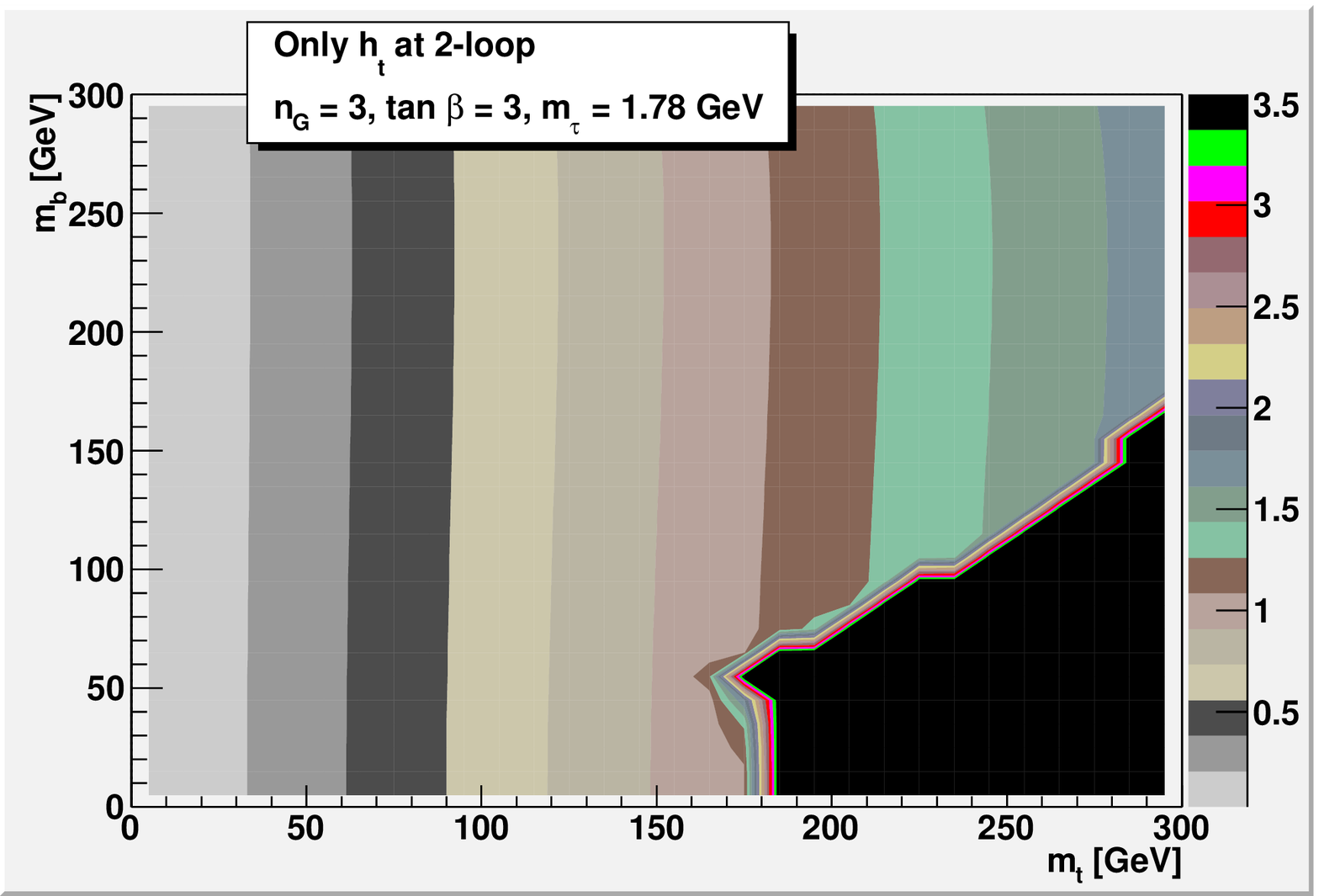}}
\subfigure{\includegraphics[width=0.45\textwidth]{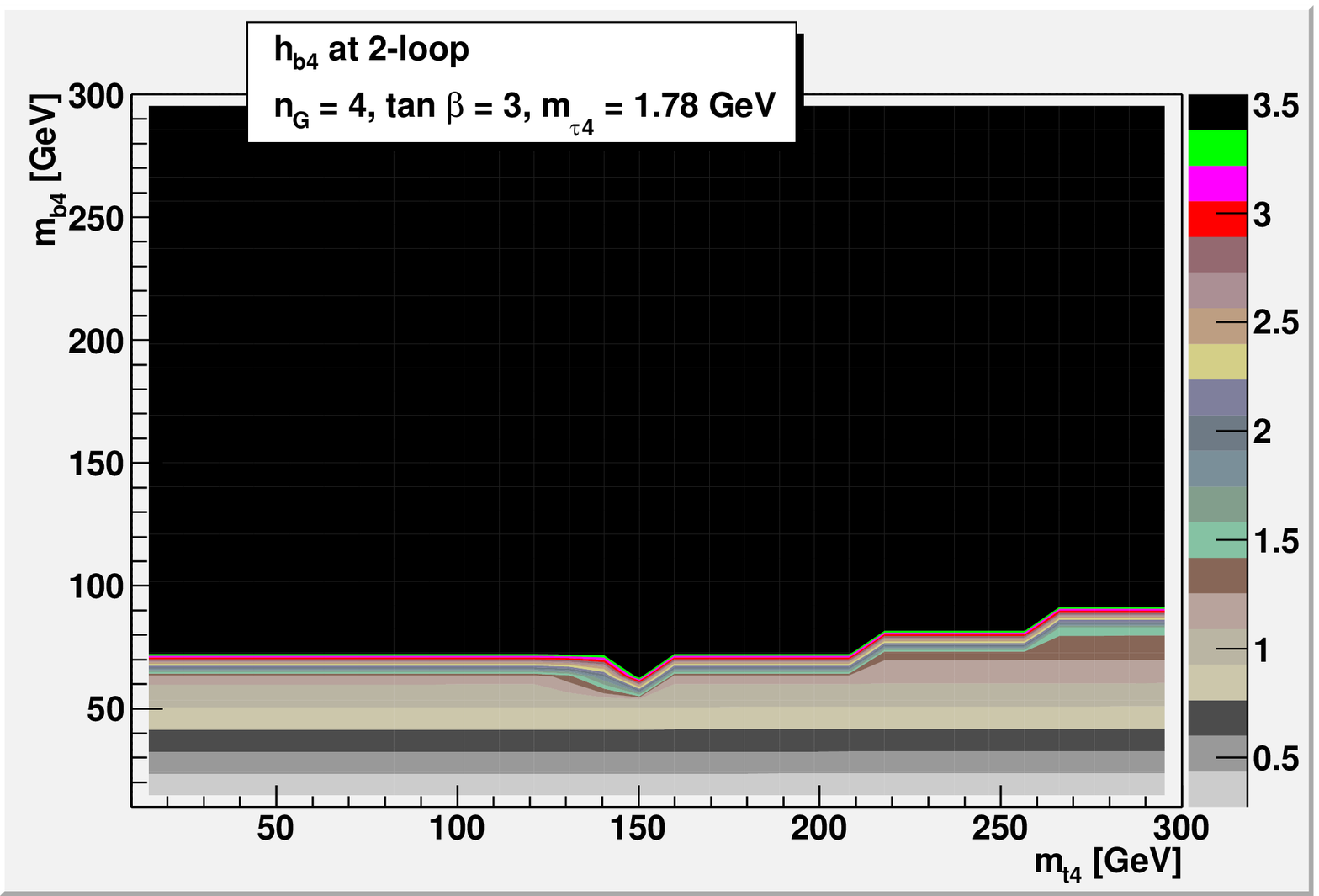}}
\subfigure{\includegraphics[width=0.45\textwidth]{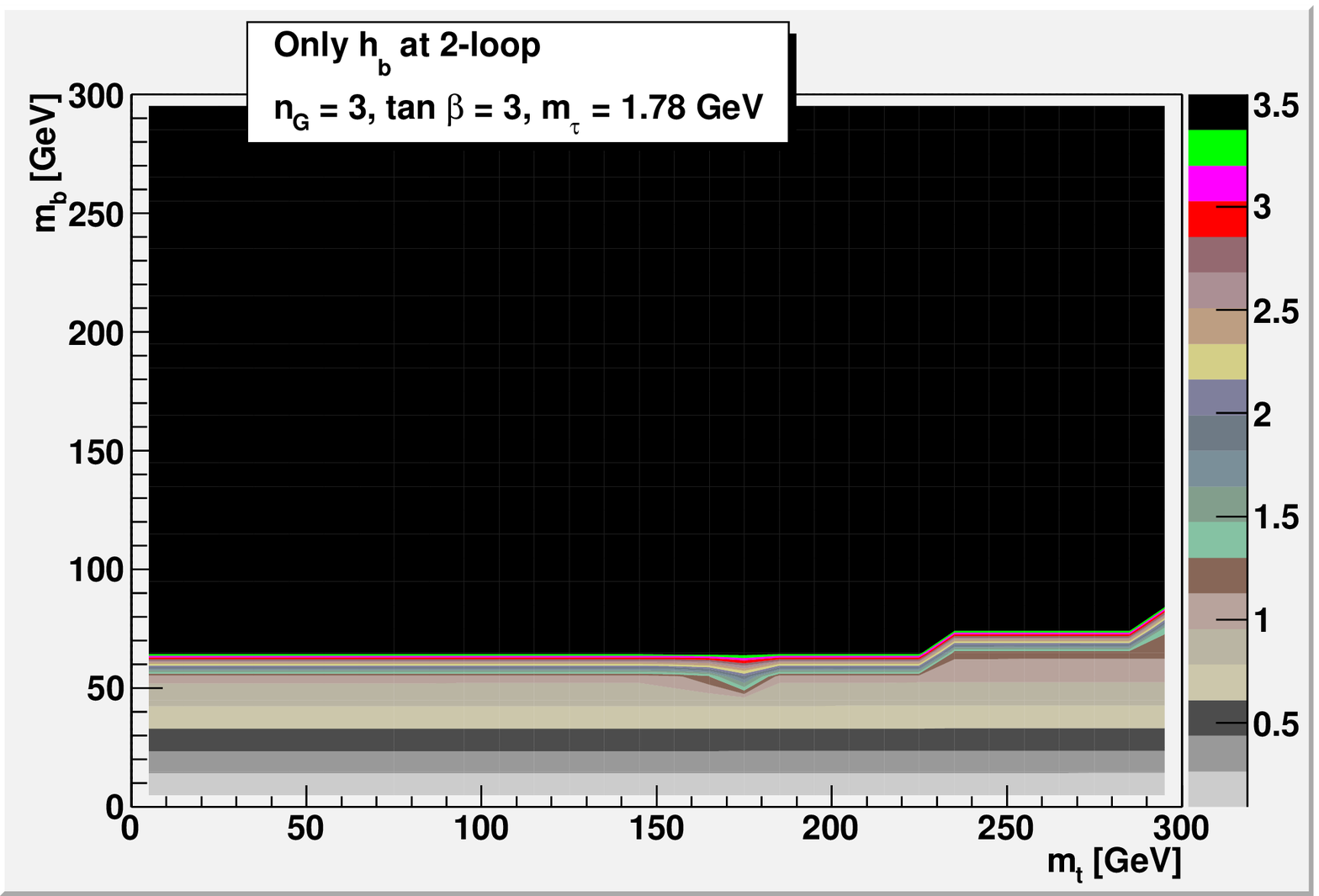}}
\subfigure{\includegraphics[width=0.45\textwidth]{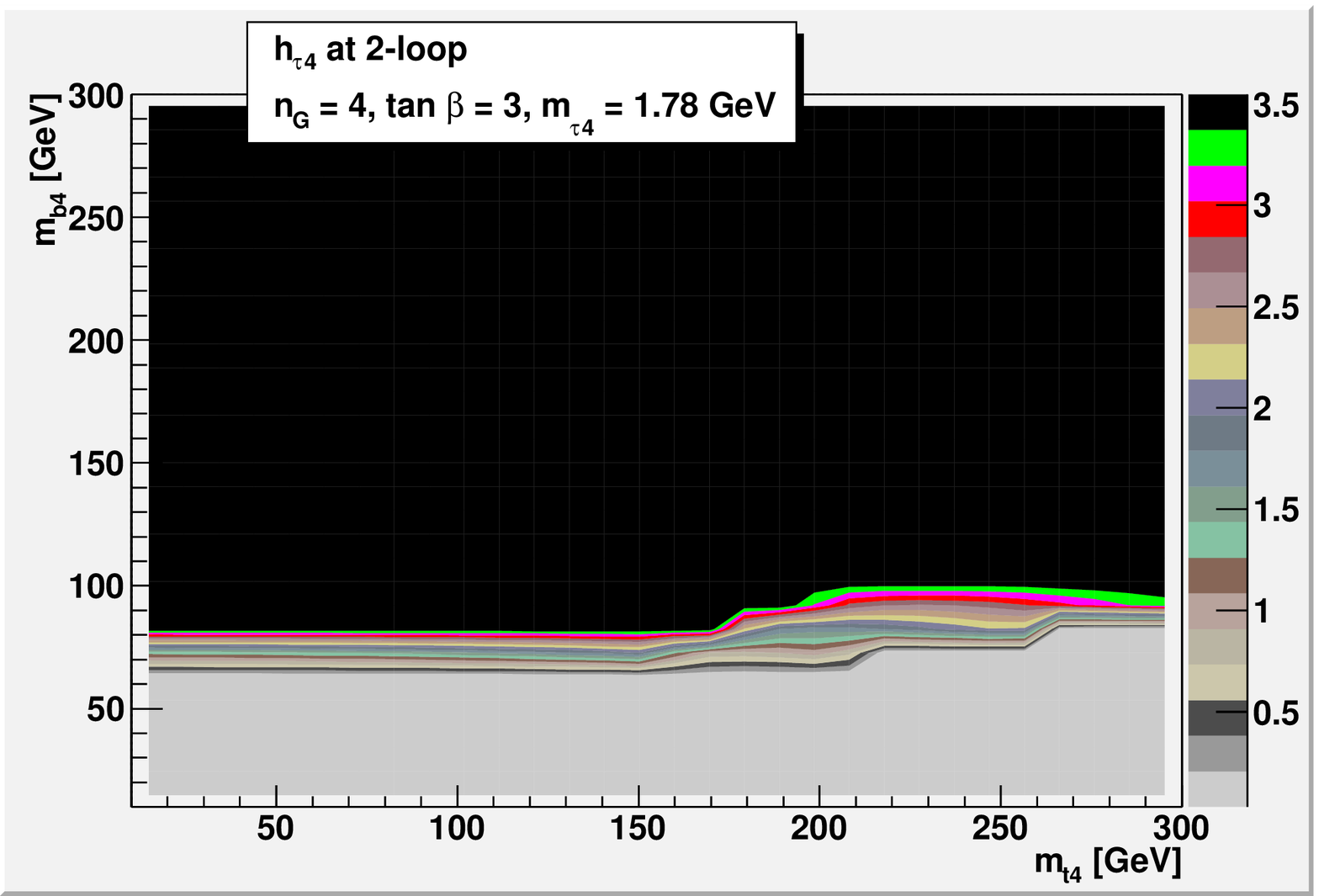}}
\subfigure{\includegraphics[width=0.45\textwidth]{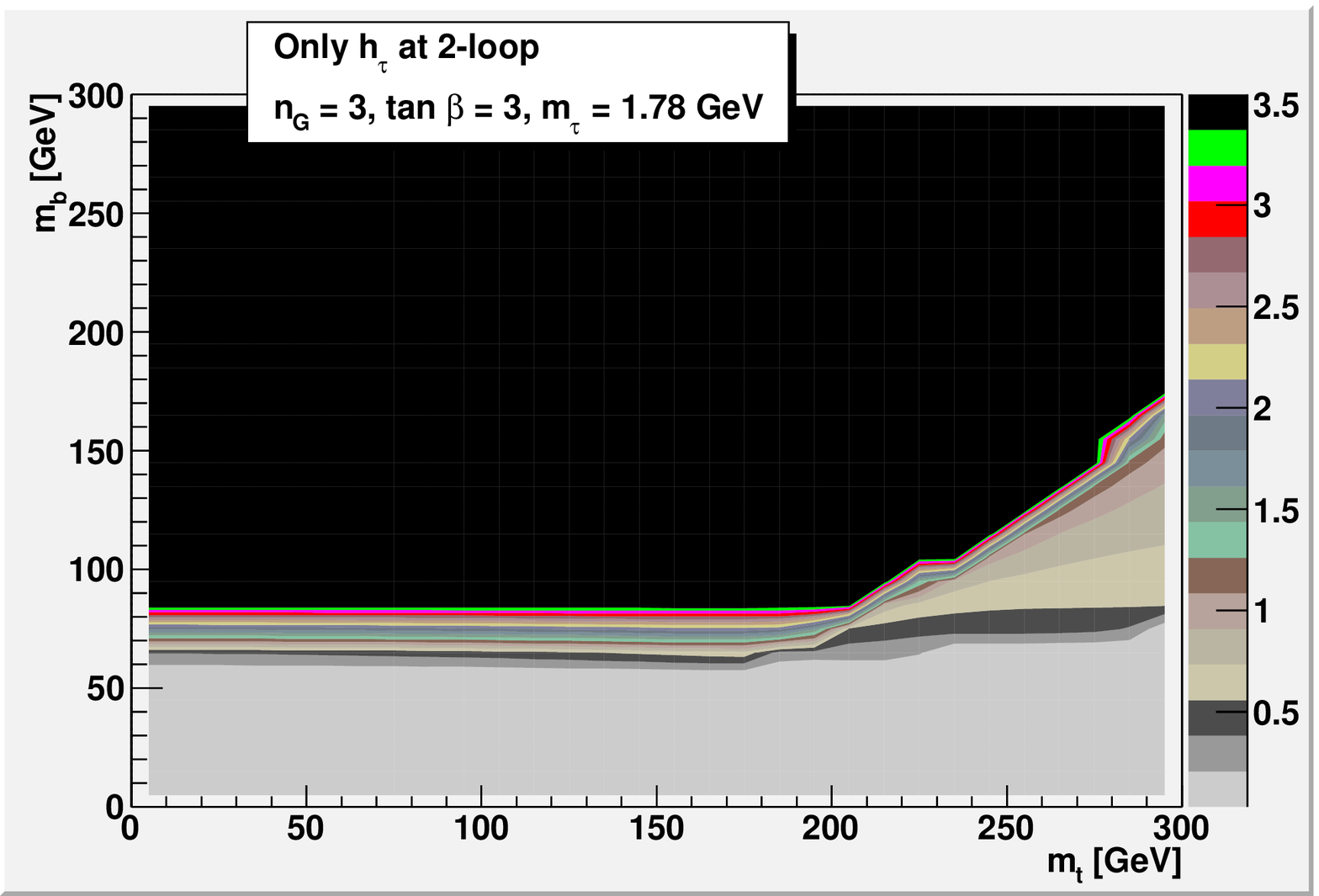}}
\subfigure{\includegraphics[width=0.45\textwidth]{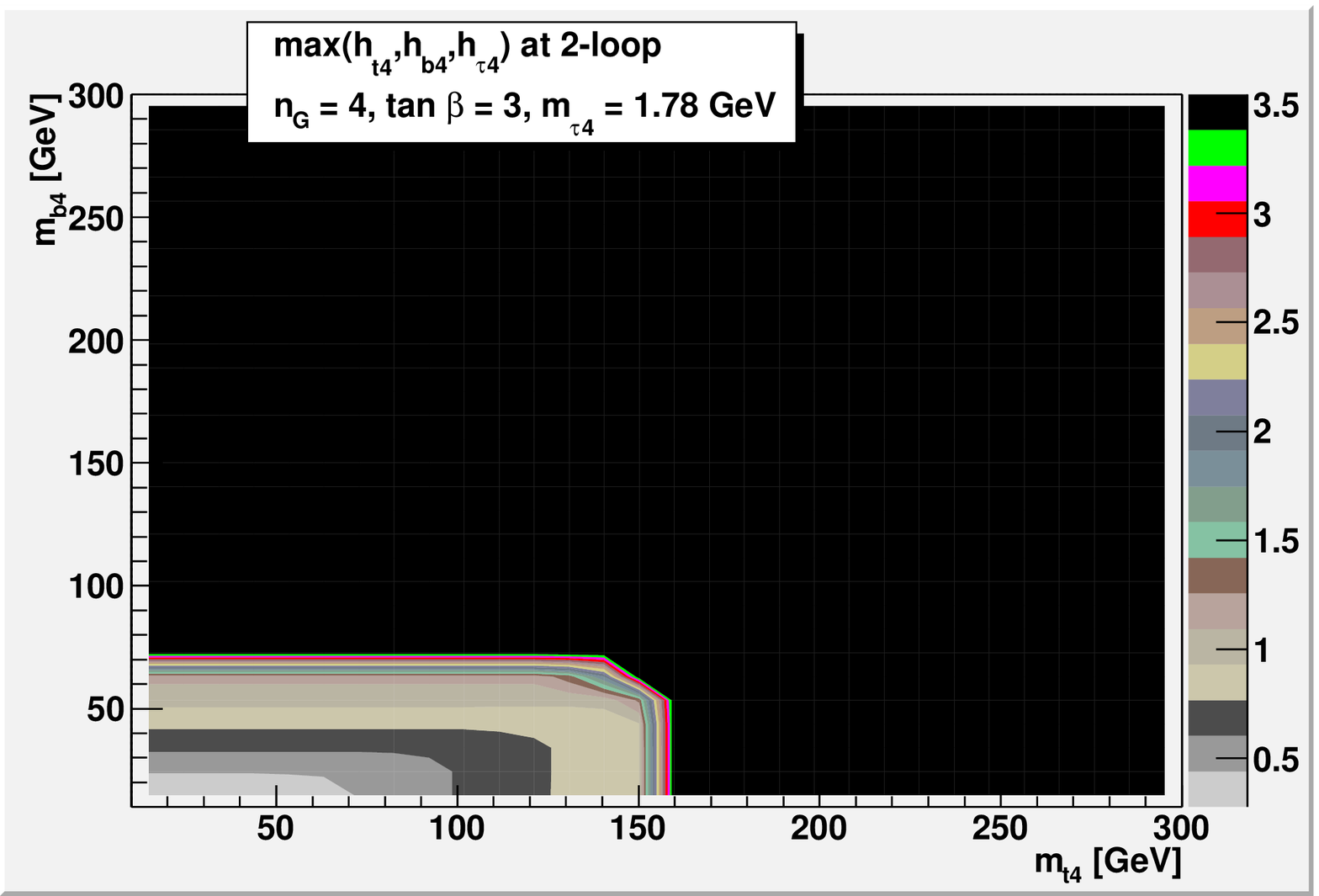}}
\subfigure{\includegraphics[width=0.45\textwidth]{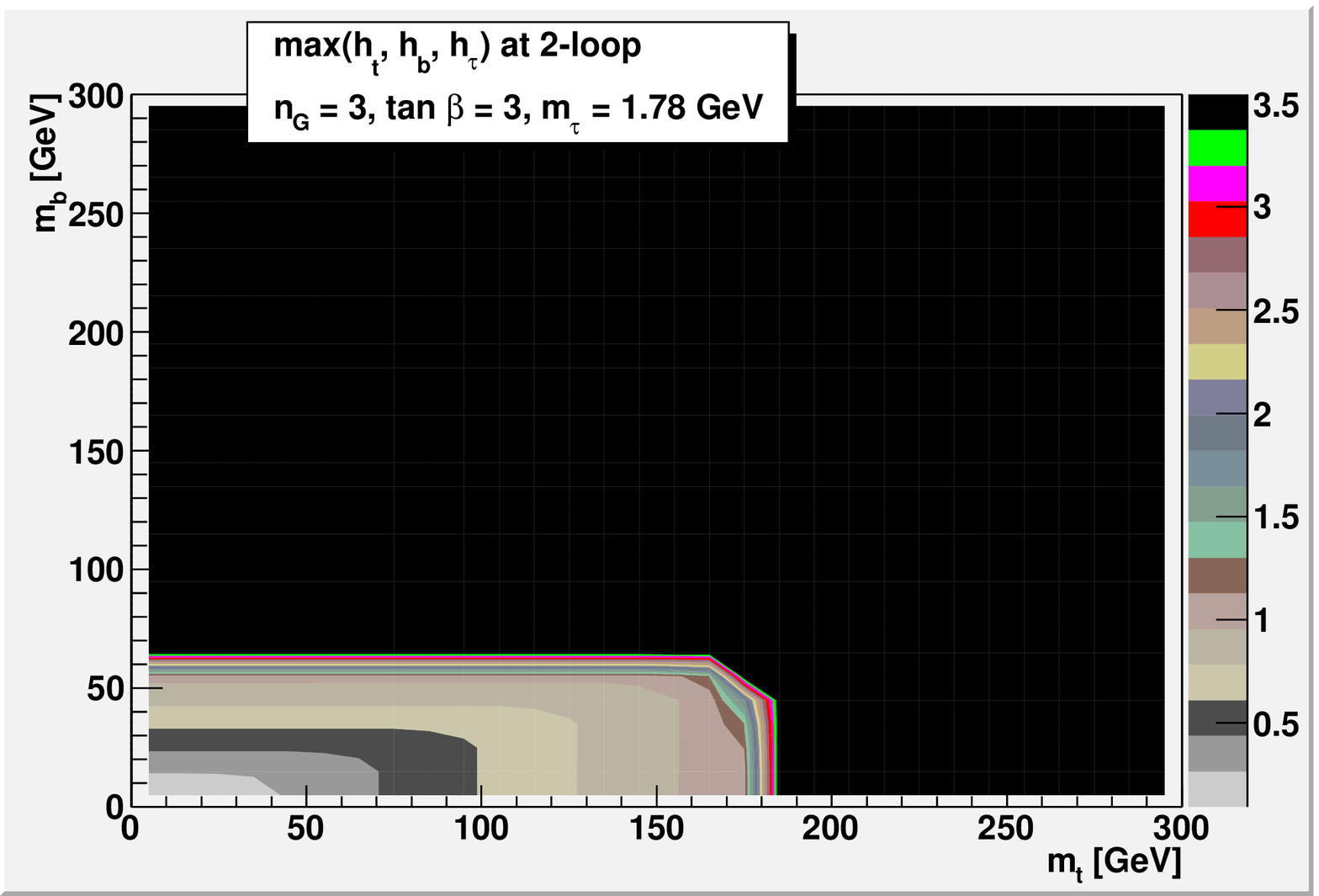}}
\caption{\footnotesize The first three rows of the left and right panels show $h_{t'}$, $h_{b'}$, $h_{\tau'}$ in MSSM4, and $h_{t}$, $h_{b}$, $h_{\tau}$ in MSSM3, respectively, at 2-loop and
$M_X= 2.3 \times 10^{16}$ GeV, plotted in the
$m_{t'}$--$m_{b'}$ and $m_{t}$--$m_{b}$ plane for $\tan \beta = 3$ and $m_{\tau'}=m_{\tau}=1.78$ GeV. For the MSSM4 case this experimentally excluded
value is chosen for illustrative purposes only. The black regions
indicate where the Yukawa couplings become non-perturbative. The last
row shows the regions where all Yukawa couplings are perturbative.}
\label{mssmfourtaupeqtau}
\end{figure}

\medskip

In the left column of \vref{mssmfourtaupeqtau}, we plot the regions in the $m_{b'}-m_{t'}$ plane where the Yukawa couplings  remain perturbative all
the way up to the GUT scale using 2-loop MSSM4 RGE. We have chosen 
$\tan\beta = 3$.  In the first row, the condition that only the top-prime 
Yukawa coupling remains perturbative all the way is plotted.  
The perturbativity limit is taken to be $\sqrt{4 \pi} \simeq 3.54$.  
All the  regions are colour-coded with the values that the Yukawa couplings 
take at the high scale. The legend for the colour-coding is shown on the 
right side of each plot.  If the Yukawa coupling attains a value beyond the
upper limit of $3.54$, the point is flagged as non-perturbative and is denoted 
in black. Thus, in each of these plots regions in black are ruled out by the
perturbativity limit.  From this first plot of the left column, we see that a 
large region in $m_{t'}$ opens up  as  the $b'$ mass increases. For  
$m_{b'} >  70$ GeV, $m_{t'} > 150$ also can be made valid. However, although 
 the $t'$ Yukawa coupling is perturbative in these regions, the other two Yukawa 
couplings are not, as is evident from the plots in the next two rows.  In the 
second row, left column, we exhibit the same plots, however requiring that 
only $h_{b'}$ remains perturbative, and in the third row, left column, 
requiring that only $h_{\tau'}$ remains perturbative.  As we see from the plots,  
although the $t'$ Yukawa coupling  is perturbative, the $b'$ and 
$\tau'$ Yukawa couplings are no longer perturbative in these regions. The last row shows  
plots where all the constraints are put together.  Here we see that for a 
negligible $\tau'$ mass ($\sim $ 1.75 GeV), $t'$ and $b'$ masses are 
constrained to be:
\begin{equation}
\label{massrangemtaupeqmtau}
0  \lesssim m_{b'} \lesssim 70 \text{ GeV}, \quad  0  \lesssim m_{t'} \lesssim 160 \text{ GeV}, \quad  m_{\tau'} = 1.78 \text{ GeV}, \quad \tan\beta = 3. 
\end{equation}

It is instructive to compare the above results with those of the MSSM3.  
In MSSM3, as we know, the Yukawa couplings do remain perturbative all the 
way up to the GUT scale for the known masses of the top and  bottom quarks and the 
tau lepton, the only exception being in regions where  $\tan\beta$ is very 
small or very large.  In the right panel of  \ref{mssmfourtaupeqtau},
we present the analogous analysis for the case of the MSSM3. We have fixed 
the $\tau$ mass at its experimental value but varied $m_{t}$ and $m_{b}$. 
We have further fixed $\tan\beta$ to be 3 as in the case of four generations. 
The results in the case of MSSM3 are strikingly  similar to those obtained for
the  four generational MSSM4.  In fact  we can read off from the plot
the regions in $m_{t} - m_{b}$ plane where all the three Yukawa couplings 
remain perturbative:
\begin{equation}
0  \lesssim m_{b} \lesssim 60 \text{ GeV}, \quad  0  \lesssim m_{t} \lesssim 180 \text{ GeV}, \quad \tan\beta = 3. 
\end{equation}
The reason for these similar results is the way Landau poles appear in the Yukawa couplings. The evolution of the
Yukawa couplings depends only on themselves and the gauge couplings. As it is clear from the RGEs (see Appendix~\ref{appendix-upperbounds}), the evolution of the
gauge couplings is very similar to that of the three generation case,  except for the $\beta$-functions $b_i$ which change 
the slope of the $h(t)$. The change in the $b_i$ for the three vs.~four generations is not very large to induce large changes
in the upper limits. The difference is within a few tens of GeV. 

\begin{figure}[h!]
\centering
\subfigure[Constraints in the $m_{b'}$--$m_{t'}$ plane  from the perturbativity of 
$h_{t'}$ for fixed values of $m_{\tau'}=100.8$ GeV and $\tan \beta=3$.]{\label{mssm4yukawamtaupexp}  {\includegraphics[width=0.45\textwidth]{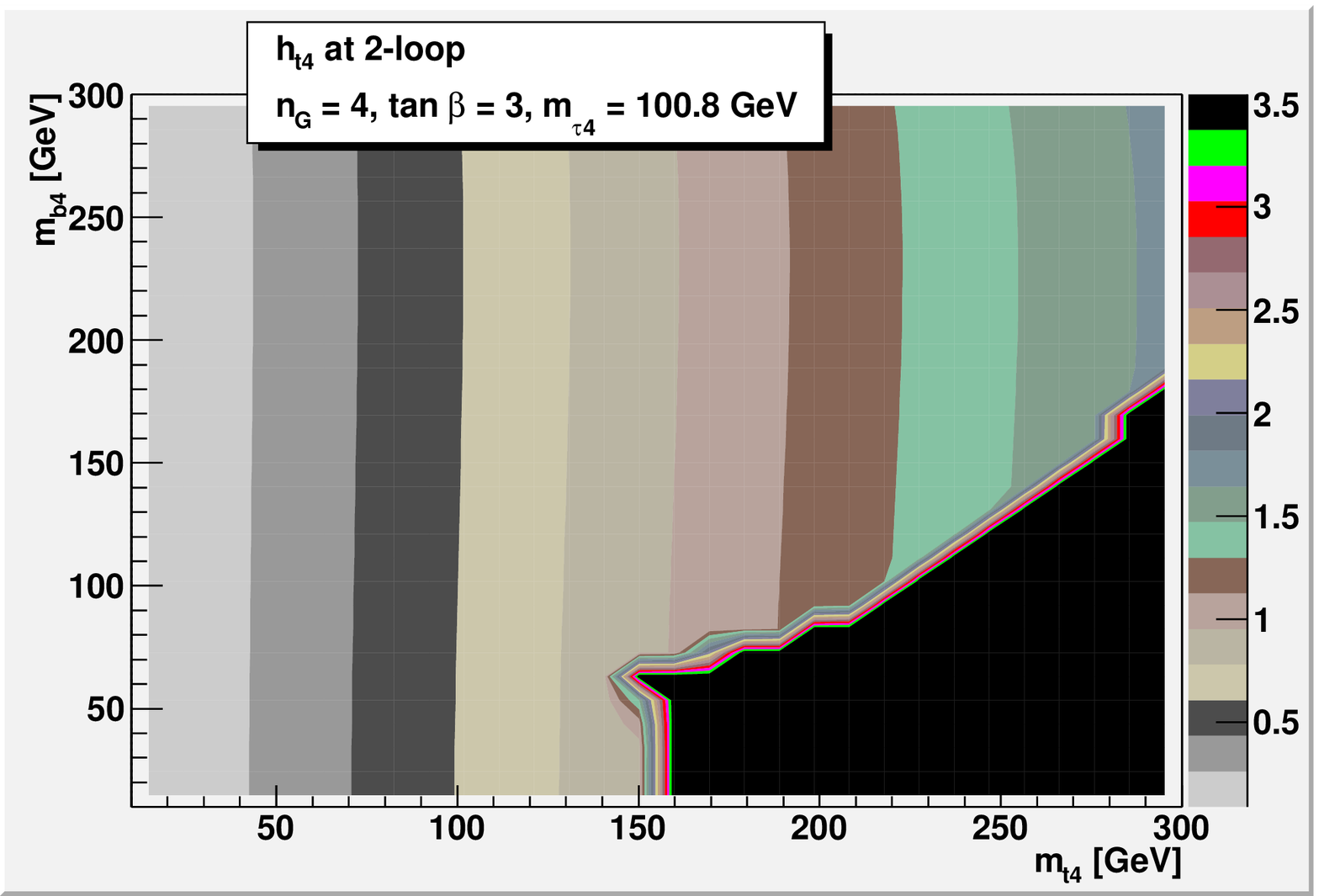}} }
\subfigure[Constraints in the $m_{b'}$--$m_{\tau '}$ plane from the perturbativity
of the max $(h_{t'}, h_{b'} , h_{\tau '}$), for fixed values of $m_{t'}=150$ GeV 
and $\tan \beta=3$.]
{ \label{mssm4yukawafixedmtp} {\includegraphics[width=0.45\textwidth]{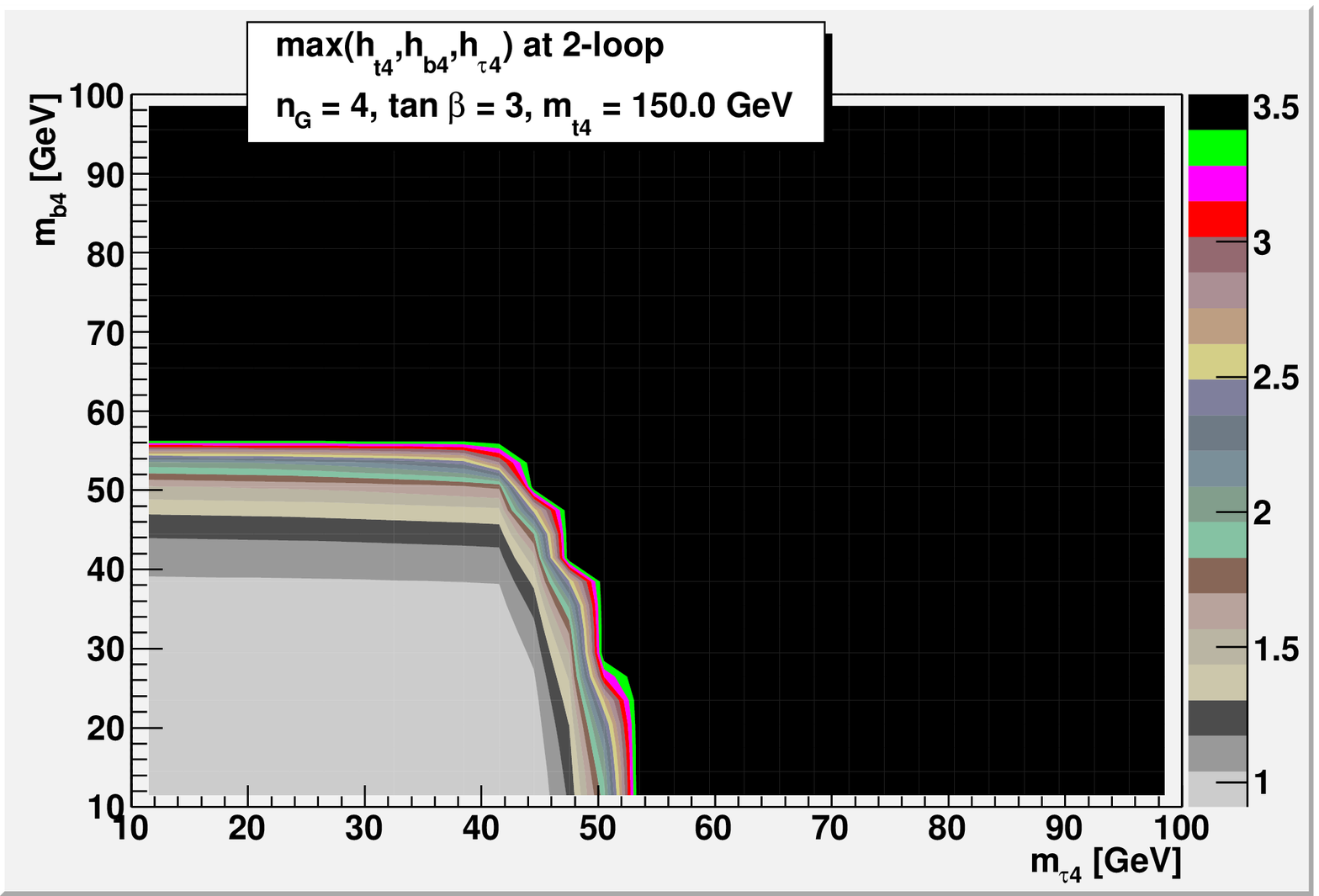}} }
\caption{\footnotesize MSSM with 4 generations at 2-loop. The black region is where the respective Yukawa couplings become non-perturbative below the unification scale $M_X = 2.3\times10^{16}$ GeV.}
\label{pertfig2}
\end{figure}

The results of \vref{massrangemtaupeqmtau} are valid only when $m_{\tau'} = m_{\tau}$.  Increasing the value of $m_{\tau'}$ to the lower limit
of \vref{pdglimits} significantly modifies these results. We find that as expected $m_{t'}$ which is less dependent on the $\tau'$ mass remains 
perturbative for a similarly large region of the parameter space as shown in \ref{pertfig2}\protect\subref{mssm4yukawamtaupexp}. However, for the values,
\begin{equation}
\label{massrangemtaupeq100}
0  \lesssim m_{b'} \lesssim 0 ~\text{ GeV}, \quad 0  \lesssim m_{t'} \lesssim 160 ~\text{ GeV}, \quad m_{\tau'} = 100.8 ~\text{ GeV}, \quad \tan\beta = 3,
\end{equation} 
$h_{b'}$ is not perturbative for any $m_{b'}$. In fact, $h_{\tau'}$ itself is no longer perturbative.
 
\medskip

We now proceed to keeping $m_{t'}$ fixed while varying $m_{b'}$ and $m_{\tau'}$. We have chosen two values of $m_{t'}$, the one
given by the approximate perturbative upper limit $\sim 150$ GeV (\vref{massrangemtaupeqmtau}) and the other the
experimental lower limit of 256 GeV.  In \ref{pertfig2}\protect\subref{mssm4yukawafixedmtp}, we have plotted the regions in the $m_{b'}- m_{\tau'}$ 
plane in which the theory remains perturbative all the way up to the GUT scale. From the figure we can read off the valid mass 
ranges as: 
\begin{equation}
0  \lesssim m_{b'} \lesssim 60~\text{ GeV}, \quad  0  \lesssim m_{\tau'} \lesssim 50~ \text{ GeV}, \quad \tan\beta = 3, \quad m_{t'} = 150 \text{ GeV}
\end{equation}
\begin{equation}
0  \lesssim m_{b'} \lesssim 56 \text{ GeV}, \quad  0  \lesssim m_{\tau'} \lesssim  53 \text{ GeV}, \quad \tan\beta = 3, \quad m_{t'} = 256 \text{ GeV}
\end{equation}

\begin{figure}[p]
\centering
\subfigure[For the masses in \vref{eq:experimental_limits_on_masses} \textit{without} $T$-parameter constraints.]{\label{mxtanbetaexplimits}
{\includegraphics[width=0.45\textwidth]{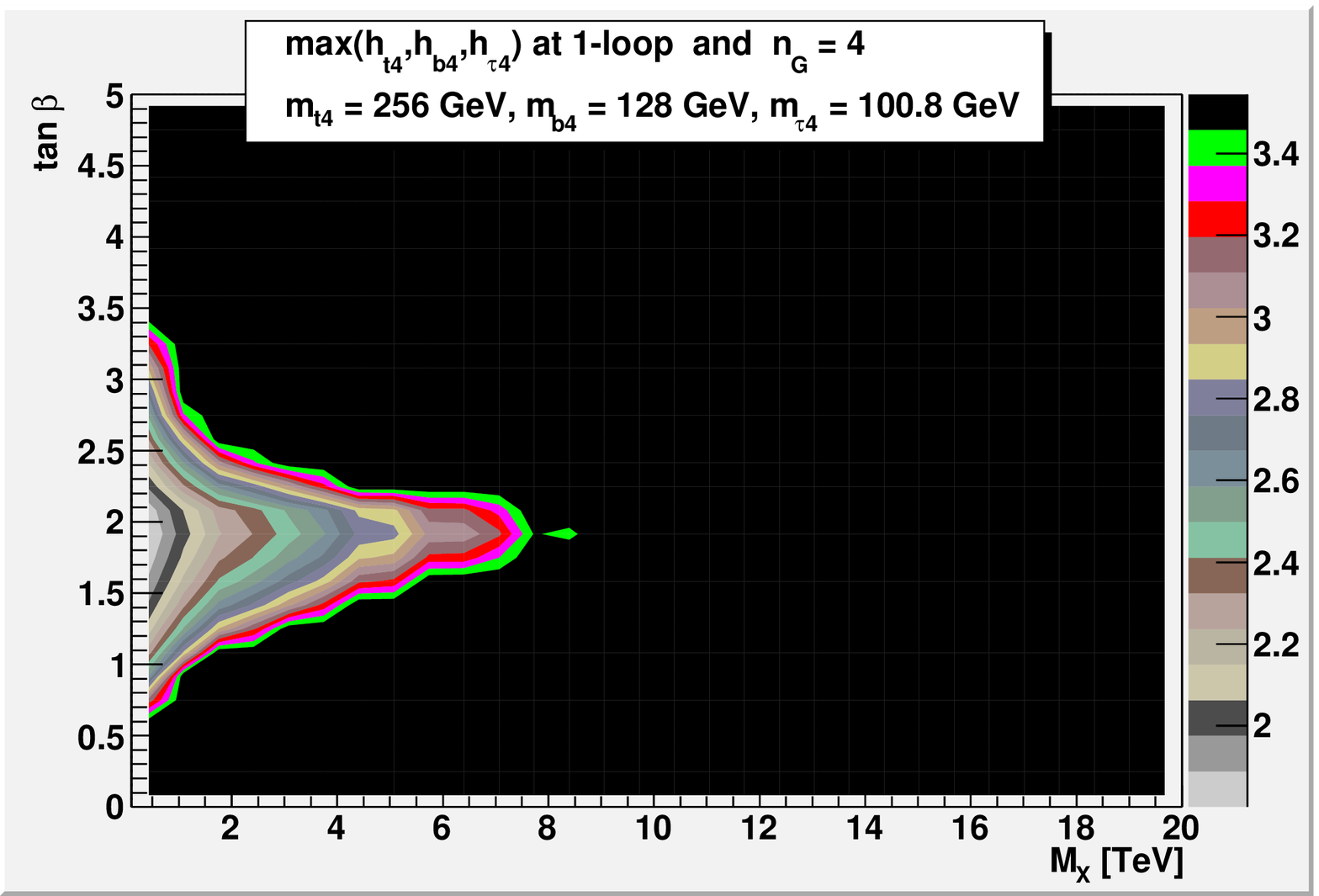}
\includegraphics[width=0.45\textwidth]{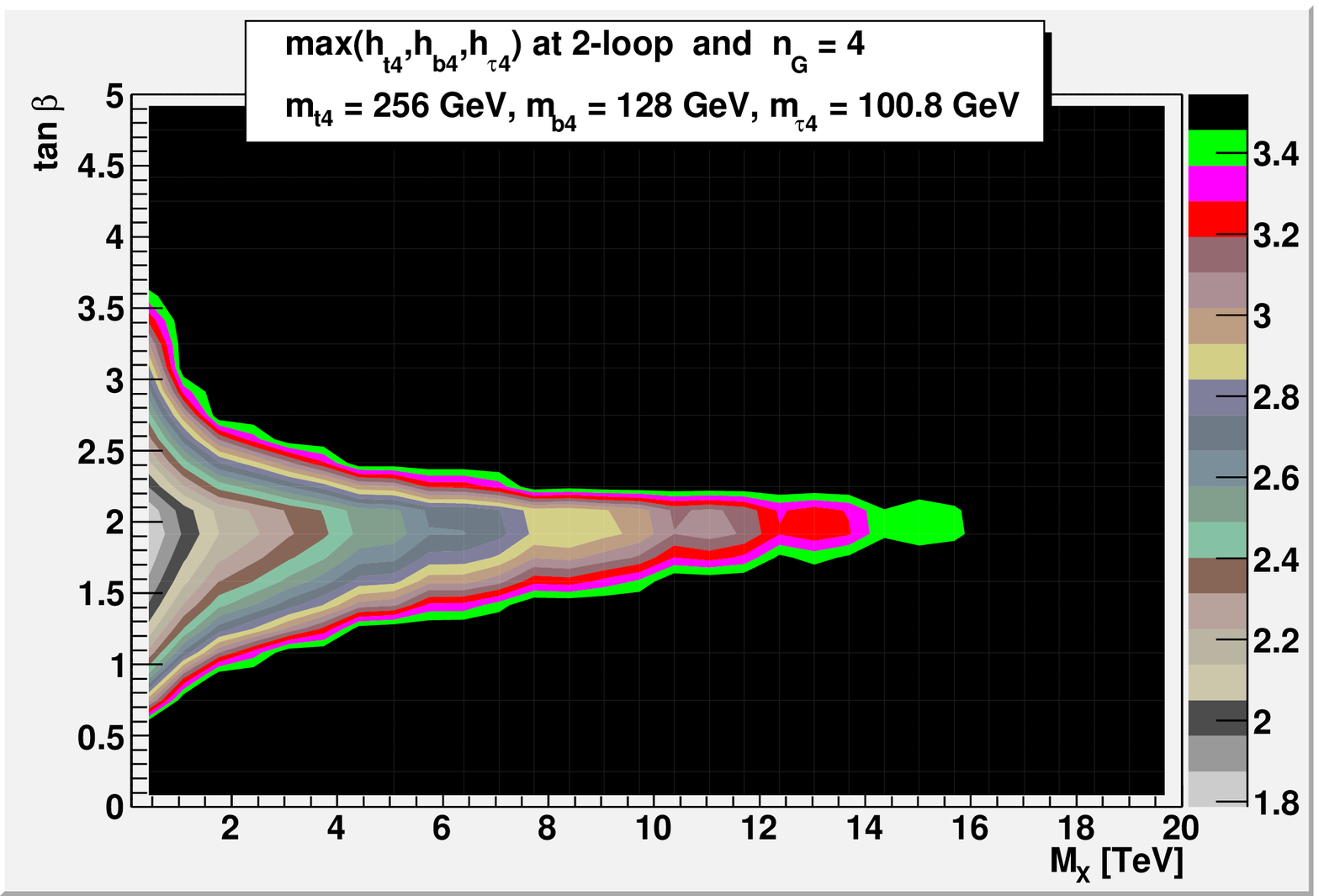}} }
\subfigure[For the masses considered in Ref.~\cite{Murdock:2008rx}.]{\label{mxtanbetanandi}
{\includegraphics[width=0.45\textwidth]{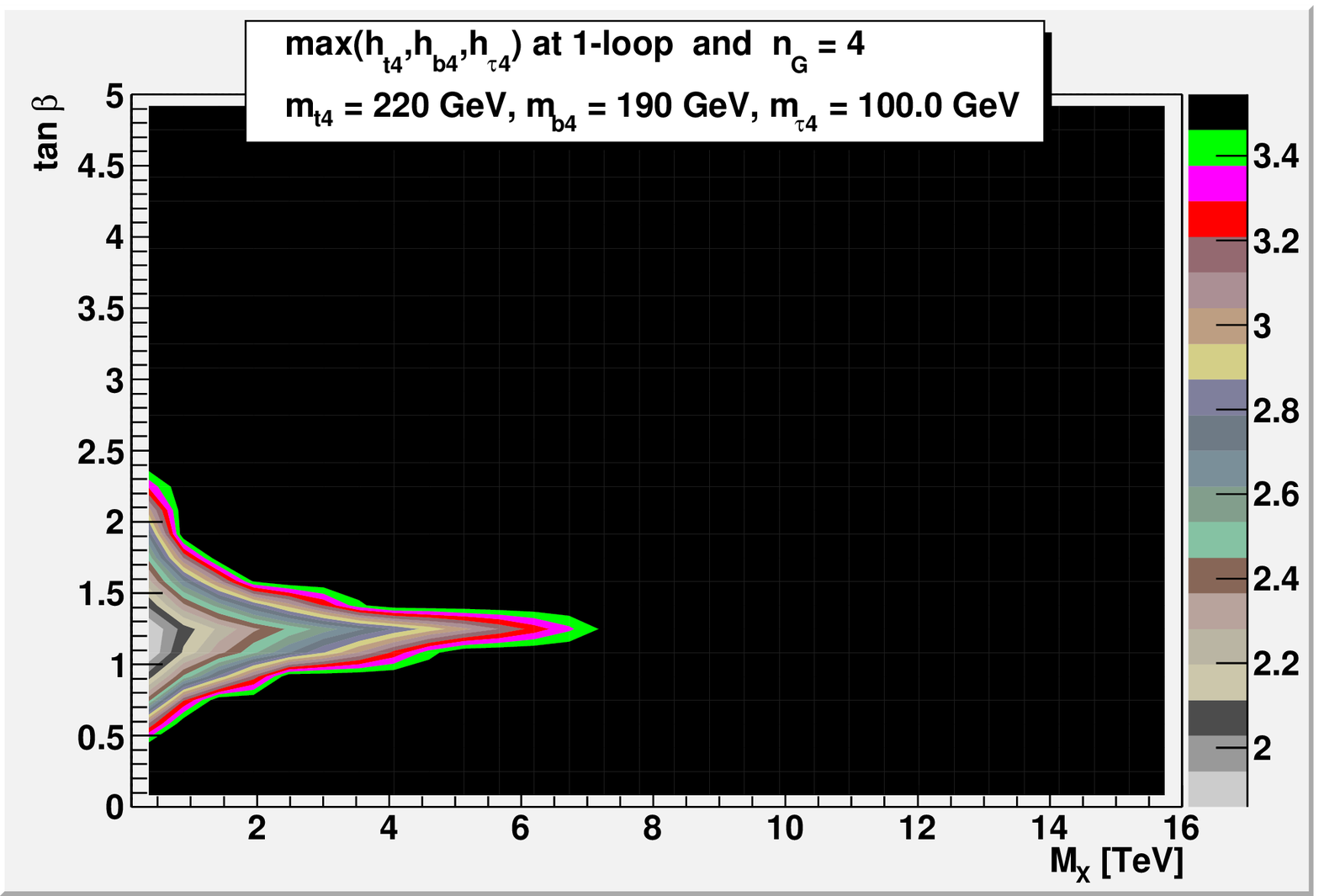}
\includegraphics[width=0.45\textwidth]{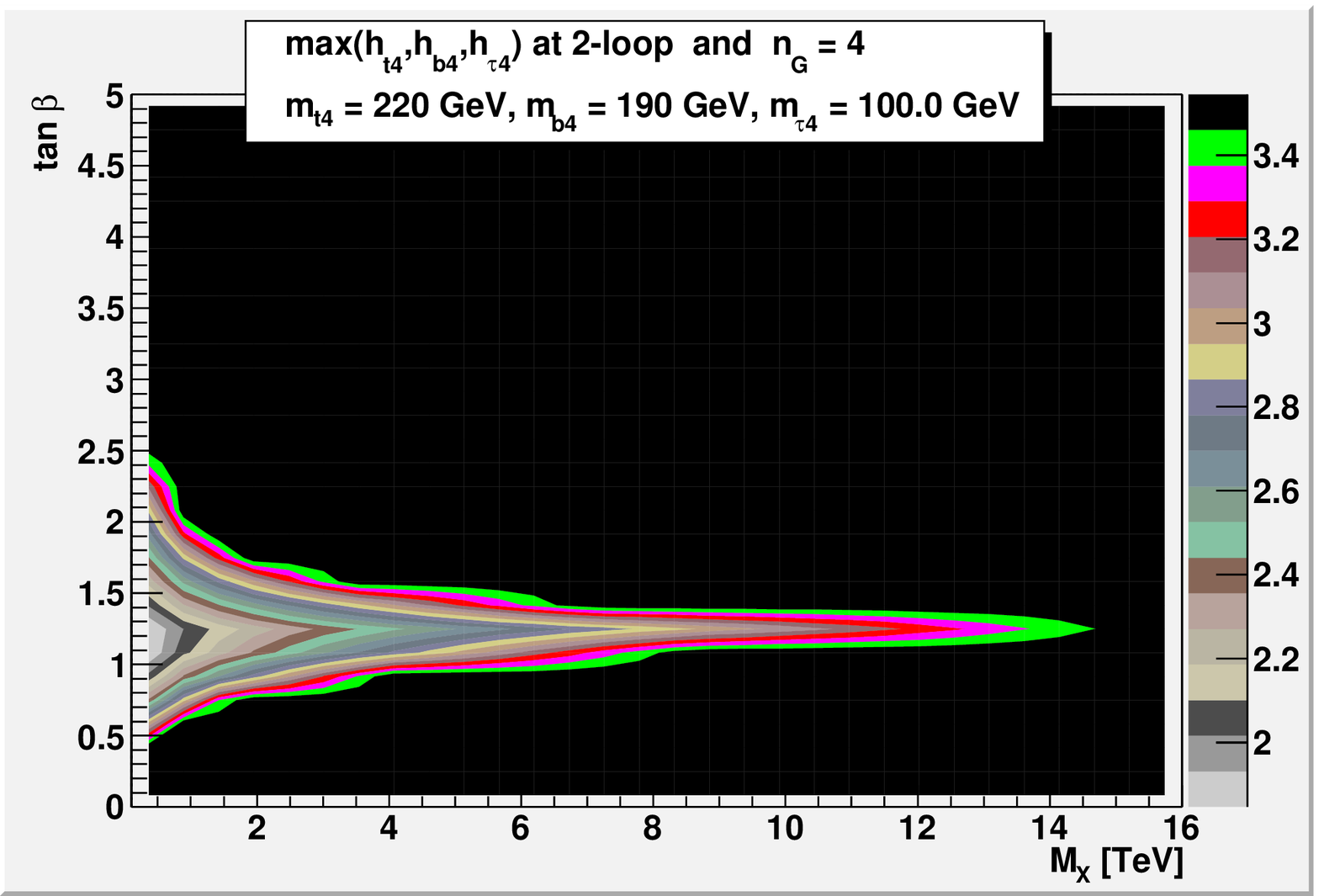}}}
\subfigure[For the masses in \vref{ourlimits} \textit{with} $T$-parameter constraints.]{\label{fig:mxtanbetaourlimits}
{\includegraphics[width=0.45\textwidth]{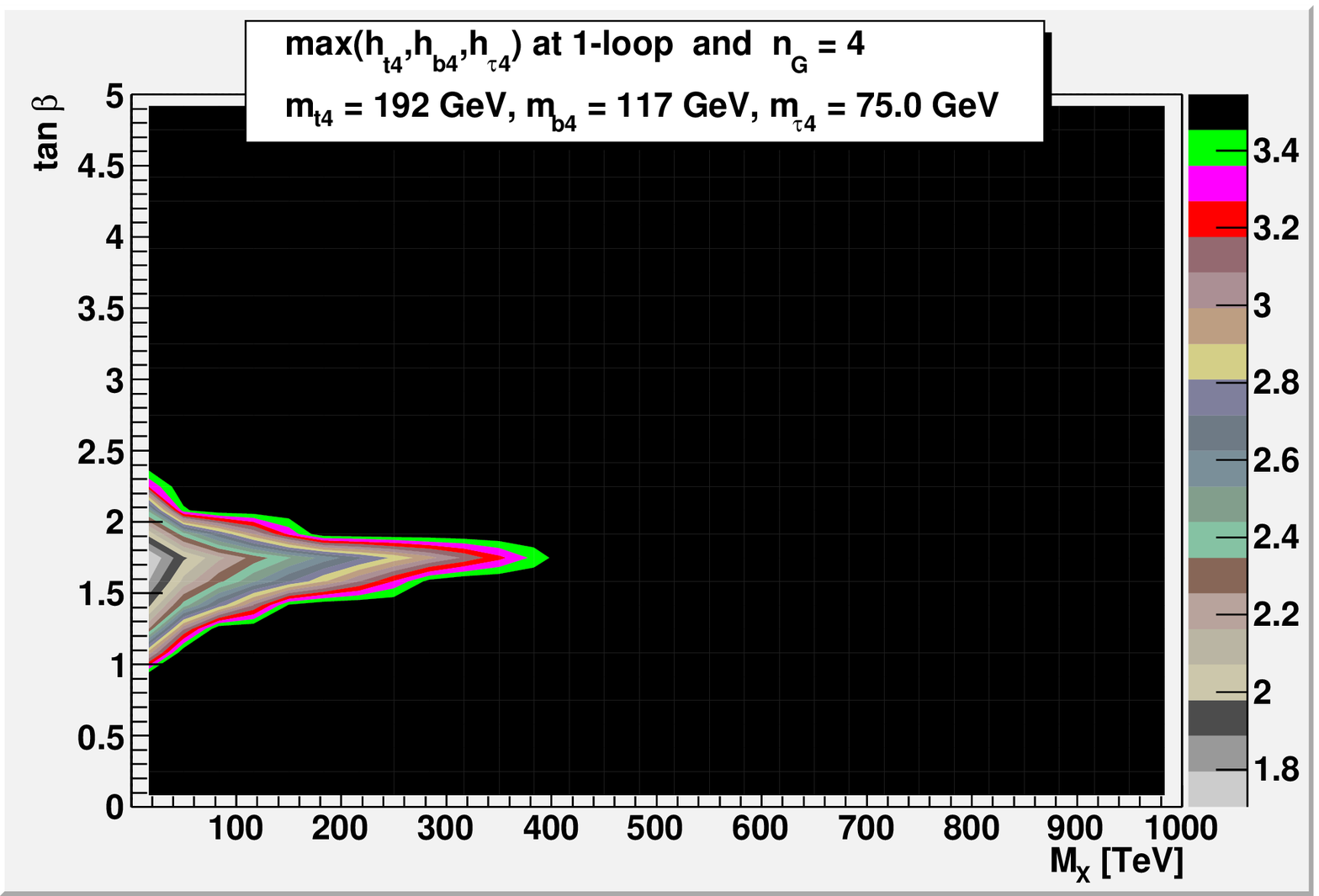}
\includegraphics[width=0.45\textwidth]{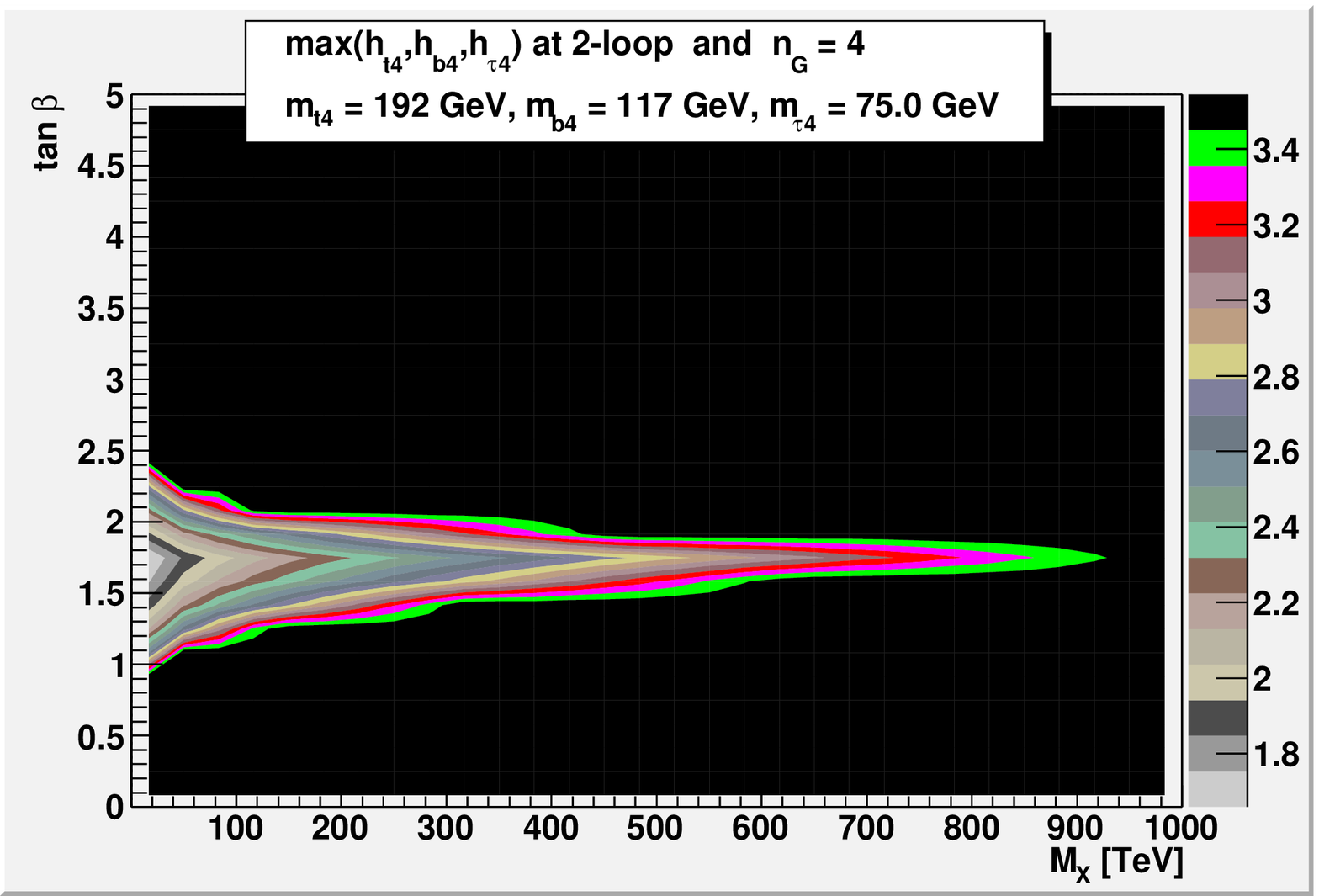}}}
\caption{\footnotesize Regions in the $M_X$--$\tan \beta$ plane where all the Yukawa couplings of the 
charged fourth generation fermions are perturbative at 1-loop and 2-loops
respectively. The different rows correspond to different values of the
fermions indicated on the panels and explained in the subcaption of 
the figure in each row.}
\label{mxtanbeta}
\end{figure}

Clearly these mass ranges are excluded by the constraints of \vref{eq:experimental_limits_on_masses} or even by the weaker constraints in \ref{pdglimits} and \ref{ourlimits} on page \pageref{ourlimits}. 

\medskip

To restore the perturbativity in the theory we can take either of the two approaches (i) add new particles
with new Yukawa contributions so as the keep the Yukawa couplings perturbative all the way up to the
GUT scale  (ii) take the view point that the theory is valid only up to a scale which is allowed by 
perturbative constraints and then some new non-perturbative physics takes over.  We will study the
second option in the present work and note that the first option has already been considered by others. 
To this end, we do a complete scan of the allowed regions of the high scale, $M_X$, and $\tan\beta$ for  a given set of
values of $m_{t'}, m_{b'}, m_{\tau'}$.  In \vref{mxtanbeta}, we present the results using  the lower limits on the fourth
generation masses 
given in \ref{pdglimits} and \ref{ourlimits} on page \vpageref{ourlimits}.  We have demanded that all the Yukawa couplings remain perturbative up to 
the scale $M_X$. We have presented the results using both 1-loop as well as 2-loop RGE to show the importance
of using 2-loop RGE. From the 1-loop plot we see that for $\tan\beta\sim2$,  the Yukawa couplings barely
remain perturbative up to 8 TeV or so. At the two loop level this scale increases to about  16 TeV.  

\medskip

We have also compared our analysis with that of
Ref.~\cite{Murdock:2008rx}. We have chosen the same masses $m_{t'} =
220$ GeV, $m_{b'} = 190$ GeV and $m_{\tau'} = 100$ GeV. The allowed
regions in $M_{X}$ vs.~$\tan\beta$ plane are presented in
\ref{mxtanbeta}\protect\subref{mxtanbetanandi} on page
\vpageref{mxtanbeta}. We find that while qualitatively we agree with them, quantitatively we differ in the maximal value of $M_X$ by a couple of orders in magnitude.

\medskip

\ref{mxtanbeta}\protect\subref{fig:mxtanbetaourlimits} on page \vpageref{mxtanbeta} shows that if we choose the fourth generation fermion
masses to be at the weaker lower limits (see \vref{ourlimits}), we can have perturbativity
all the way up to $\sim$ 1000 TeV. Note that these weaker lower limits are 
obtained under very reasonable assumptions.
Finally,  the  relevant $M_X$ and $\tan\beta$ ranges can be read out 
from the plots as follows : 
\begin{align}
1 \lesssim \tan\beta \lesssim 3.6, \quad 1 \text{ TeV}  \lesssim M_X &\lesssim 16  \text{ \enspace TeV} \quad\text{for \ref{pdglimits} on page \pageref{pdglimits}}\nonumber \\
1 \lesssim \tan\beta \lesssim 2.4, \quad 1 \text{ TeV}  \lesssim M_X &\lesssim 920 \text{ TeV} \quad\text{for \ref{ourlimits} on page \pageref{ourlimits} }
\end{align}

\subsection{ Implications for mSUGRA} 
Before closing  this section, let us comment on the possibility of realising minimal supergravity with four generations. 
From a phenomenological point of view such a possibility would be interesting with supersymmetric partners of
the fourth generations leading to new experimental signatures at the weak scale. Further, due to the presence
of additional Yukawa couplings, the weak scale supersymmetric mass spectrum would most likely be quite
different from that of mSUGRA3 \cite{Gunion:1995tp}. If the fourth family Yukawa couplings are large,
they could contribute significantly to the lightest Higgs mass at the 1-loop level, thus alleviating the little hierarchy
problem \cite{Fok:2008yg}, \cite{Litsey:2009rp}. 
However, in the standard picture of mSUGRA with universal or non-universal soft masses at the GUT scale, radiative electroweak 
symmetry breaking induced by the large $t'$ (and possibly $b',\tau'$)  Yukawa coupling and the neutralino as the lightest supersymmetric
particle, would require the theory to remain perturbative all the way up to the GUT scale. From the analysis
above, we see that theory would remain perturbative with four generations only if the fourth generation masses
are much lower than their experimental lower limits, in fact, closer to the third generation masses.
In the following we will consider a toy model of mSUGRA4, where the masses of the fourth generation are set
equal to the their third generation counterparts: 
\begin{equation*}
m_{t'} = m_t, \quad m_{b'} = m_b, \quad m_{\tau'} = m_{\tau}.
\end{equation*}
We then compute the mass spectrum at the weak scale for a sample point and compare it with that of mSUGRA3. We use our new tool \indisoft{}
to compute the spectrum at the weak scale. 

\begin{figure}[h!]
\centering
\subfigure[Three generations.]{\includegraphics[width=0.45\textwidth]{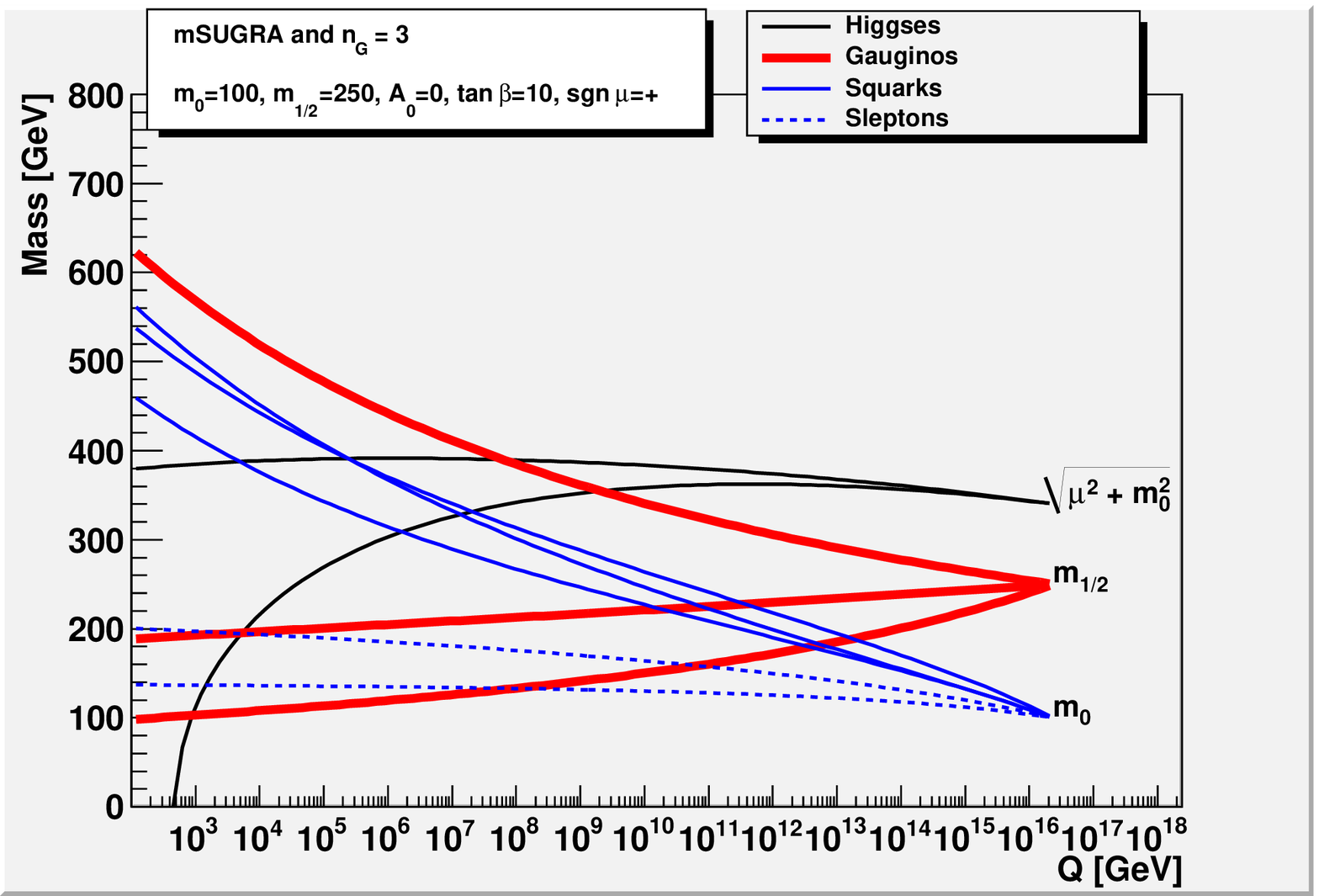}}
\subfigure[Four generations.]{\includegraphics[width=0.45\textwidth]{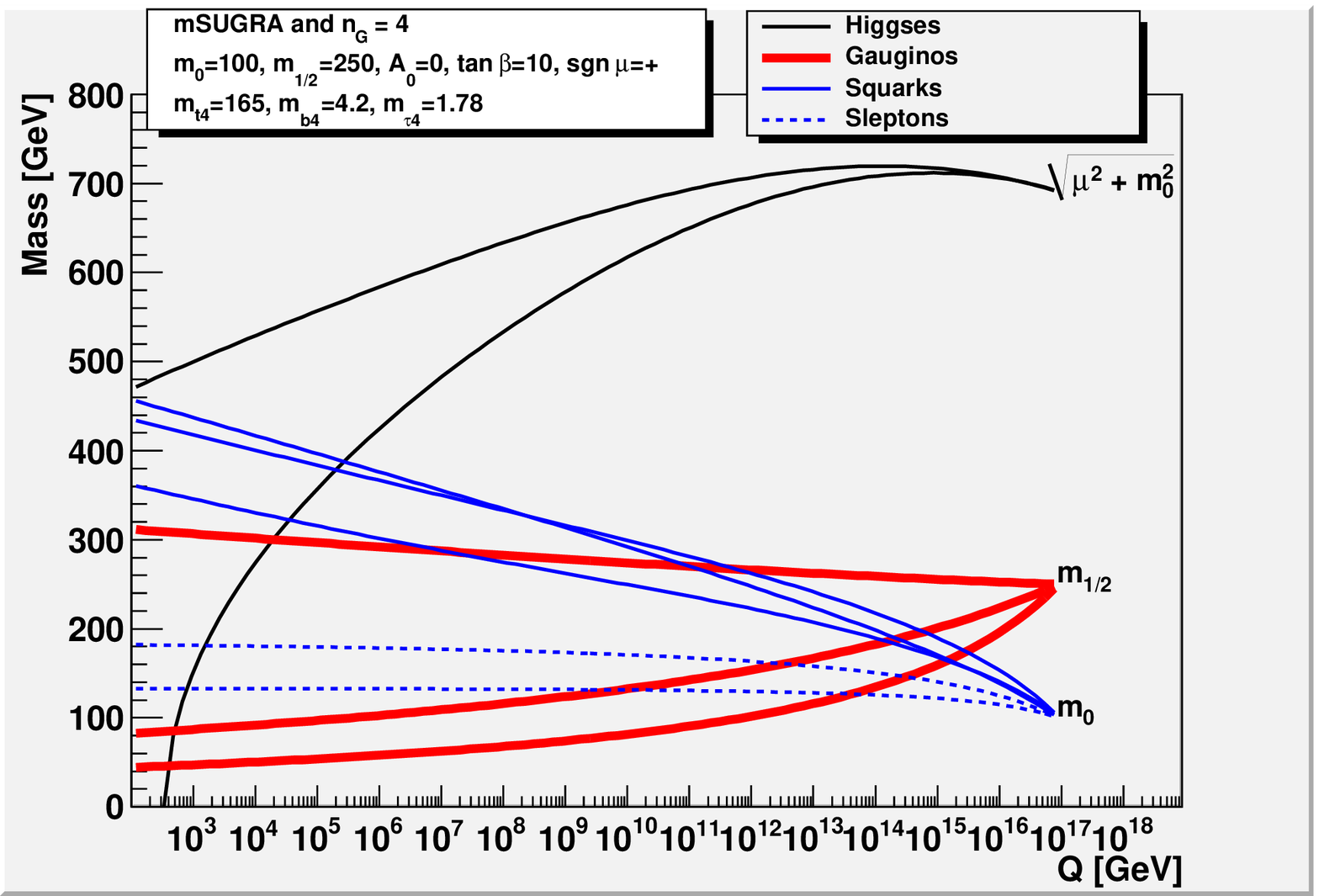}}
\caption{\footnotesize The running of the various soft masses in the MSSM3 and MSSM4 is shown in the left and right panel, respectively. The legend is indicated in the panel. The Higgses are the black lines starting at $\sqrt{\mu^2+m_0^2}$.}
\label{4gtoyrun}
\end{figure}

The very presence of four generations, irrespective of whether their Yukawa couplings are large or not, 
can lead  to interesting differences  between the three and four generation RG evolution of the soft 
masses.  In \vref{4gtoyrun} we show the running patterns of the soft terms for three
 as well as four generations.  As expected, the Higgs mass terms (black line) run to more negative
regions as compared to three generation case, and in fact, they can both become negative even for small
$\tan\beta$ in the four generation case. Perhaps the most interesting aspect is the running of the gluino
mass, which now due to a smaller $\beta$-function\footnote{Note that $b_3 = 3$ for MSSM3 and $b_3 = 1$ for MSSM4.}, almost does not evolve (up to 1-loop level) in the
four generation case (thick red line).  The sleptons (dashed-blue) are 
not significantly affected, however the squarks (undashed-blue) run to lighter values due to reduced 
gluino running effects.  All these differences in the running of the soft terms would make themselves 
evident in the mass spectrum at the weak scale.

\begin{table}[h!]
\centering
\normalsize
\renewcommand{\arraystretch}{1.4}
\begin{tabular}{|ccrc|ccrc|cr|ccl|}
\hline
\multicolumn{4}{|c|}{Higgses [GeV]} & \multicolumn{4}{c}{Gauginos [GeV]} & \multicolumn{5}{|c|}{Squarks \& Sleptons [GeV]}\\
\hline
&$h^0$   & 106.7  &&& $\widetilde{\chi}^0_1$ & 96.6  &&  $\widetilde{u}_L$   & 568.2   &&  $\widetilde{t}_1$      & 587.4\\
&$A^0$   & 382.2  &&& $\widetilde{\chi}^0_2$ & 178.3 &&  $\widetilde{u}_R$   & 547.5   &&  $\widetilde{t}_2$      & 411.0\\ 
&$H^0$   & 382.6  &&& $\widetilde{\chi}^0_3$ & 343.0 &&  $\widetilde{d}_L$   & 573.6   &&  $\widetilde{b}_1$      & 519.9\\ 
&$H^\pm$ & 390.9  &&& $\widetilde{\chi}^0_4$ & 362.8 &&  $\widetilde{d}_R$   & 546.6   &&  $\widetilde{b}_2$      & 547.2\\ 
\cline{1-4}
&        &        &&& $\widetilde{\chi}^\pm_1$ & 178.0 &&  $\widetilde{e}_L$   & 205.7   &&  $\widetilde{\tau}_1$   & 209.1\\ 
&        &        &&& $\widetilde{\chi}^\pm_2$ & 364.5 &&  $\widetilde{e}_R$   & 146.7   &&  $\widetilde{\tau}_2$   & 138.9\\ 
&        &        &&& $\widetilde{g}$   & 607.0 &&  $\widetilde{\nu}_e$ & 189.8   &&  $\widetilde{\nu}_\tau$ & 189.1\\ 
\hline
\end{tabular}
\caption{MSSM spectrum with 3 generations and mSUGRA boundary conditions: $m_0=100$ GeV, $m_{1/2}=250$ GeV, $A_0=0$ GeV, $\tan\beta=10$, \sgn{\mu}=+. The unification scale is $\mgut{} = 2.40\times10^{16}$ GeV.}
\label{msugra3sample}
\end{table}

\begin{table}[h!]
\centering
\normalsize
\renewcommand{\arraystretch}{1.4}
\begin{tabular}{|ccrc|ccrc|cr|cr|cr|}
\hline
\multicolumn{4}{|c|}{Higgses [GeV]} & \multicolumn{4}{c}{Gauginos [GeV]} & \multicolumn{6}{|c|}{Squarks \& Sleptons [GeV]}\\
\hline
&$h^0$   & 119.5  &&& $\widetilde{\chi}^0_1$ & 44.1   &&  $\widetilde{u}_L$   & 480.4   &  $\widetilde{t}_1$      & 499.7   &  $\widetilde{t}_1'$      & 498.8\\
&$A^0$   & 486.5  &&& $\widetilde{\chi}^0_2$ & 83.4   &&  $\widetilde{u}_R$   & 462.6   &  $\widetilde{t}_2$      & 357.8   &  $\widetilde{t}_2'$      & 356.4\\ 
&$H^0$   & 486.2  &&& $\widetilde{\chi}^0_3$ & 474.2  &&  $\widetilde{d}_L$   & 486.7   &  $\widetilde{b}_1$      & 432.4   &  $\widetilde{b}_1'$      & 428.7\\ 
&$H^\pm$ & 492.8  &&& $\widetilde{\chi}^0_4$ & 478.1  &&  $\widetilde{d}_R$   & 462.0   &  $\widetilde{b}_2$      & 465.9   &  $\widetilde{b}_2'$      & 466.2\\ 
\cline{1-4}
&        &        &&& $\widetilde{\chi}^\pm_1$ & 83.4   &&  $\widetilde{e}_L$   & 187.7   &  $\widetilde{\tau}_1$   & 196.4   &  $\widetilde{\tau}_1'$   & 196.2\\ 
&        &        &&& $\widetilde{\chi}^\pm_2$ & 481.4  &&  $\widetilde{e}_R$   & 142.0   &  $\widetilde{\tau}_2$   & 126.5   &  $\widetilde{\tau}_2'$   & 127.1\\ 
&        &        &&& $\widetilde{g}$   & 352.1  &&  $\widetilde{\nu}_e$ & 170.4   &  $\widetilde{\nu}_\tau$ & 169.6   &  $\widetilde{\nu}_\tau'$ & 169.6\\ 
\hline
\end{tabular}
\caption{MSSM spectrum with 4 generations and mSUGRA boundary conditions: $m_0=100$ GeV, $m_{1/2}=250$ GeV, $A_0=0$ GeV, $\tan\beta=10$, \sgn{\mu}=+. For the theory to be perturbative, we have chosen all 4th generation masses to be equal to their 3rd generation counterparts (toy model). The unification scale is $\mgut{} = 8.82\times10^{16}$ GeV.}
\label{msugra4sample}
\end{table}

In \vref{msugra3sample} and \vref{msugra4sample}, we present the weak scale spectrum for some sample point. Comparing the spectrum in the two tables, we find the following: (a) the Higgs
mass is heavier in mSUGRA4, in fact above the LEP limit (b) the lighter neutralinos in mSUGRA4 are lighter compared to mSUGRA3, with one close to 60 GeV (c) the squark and the gluino masses are also significantly lighter 
compared to mSUGRA3 (d) slepton masses do not have much of an impact and they seem to be close to those
in  mSUGRA3. 

\medskip

Thus the addition of a fourth chiral generation to MSSM3 seems to give rise to a lighter supersymmetric spectrum
at the weak scale with a less fine tuned Higgs mass. These results, though valid only in this particular toy model, seem
to indicate the possible features the supersymmetric spectrum would have if one could make theory perturbative. 
As discussed earlier, one possible way would be to add additional vector-like matter \cite{Murdock:2008rx}. An alternative approach, which we follow, is to lower the scale of supersymmetry breaking and ask whether
the above features are replicated. We thus look for a low scale supersymmetry mediation mechanism which is 
preferably close to the non-perturbative regime in the Yukawa couplings.


\section{GMSB with Four generations}
\label{sec:gmsb}

\subsection{Minimal Messenger Model}

Given that the four generational MSSM is barely perturbative up to few hundreds 
of TeV, supersymmetry breaking should be communicated within this energy scale
to the visible sector. Gauge mediated supersymmetry breaking (for a review, 
see Refs.~\cite{Giudice:1998bp} and \cite{Drees:2004jm}) is one such possibility where 
mediation scales can be as low as 10-20 TeV.   In the present section, we 
explore the possibility of  reconciling the minimal messenger model 
(MMM) of GMSB with the four generational MSSM.   

\medskip

The minimal messenger model has a set of chiral superfields (messengers) which transform as fundamentals 
under \SU{5}. They are coupled to a singlet field $S$ which parametrises the hidden sector supersymmetry breaking
 and whose F-component and scalar component attain  \textit{vev}s.  As a consequence, messengers attain
both supersymmetry conserving and supersymmetry breaking masses. This breaking information is then passed
on to the MSSM sector through gauge interactions.   The gauginos attain masses at the 1-loop level, given by 
\begin{equation}
M_i(X) ={ \tilde{\alpha}}_i (X) \, \Lambda \, g(x) 
\end{equation}
where $X \sim \langle S\rangle$ is the messenger mass scale (up to a
coupling constant), $\Lambda = \langle F_S\rangle/\langle S\rangle$
and $i= 1,2,3$ for the three gauge groups of the MSSM. The tilde on
the gauge couplings $\alpha_i$ denotes a division by $4 \pi$. $g(x)$ is the loop function with $x =
\langle S\rangle^2/ \langle F_S\rangle$. This function rises monotonically
between $g(0) = 1$ and $g(1)~ = 1.386.$ We chose $x = 1/2$ where $g(x)
\approx 1 $. The scalars attain their masses at the two loop level and
they are given by :
\begin{equation}
m_{\tilde{f}}^2(X) = \Lambda^2  \sum_{i = 1}^3  C_i \, \tilde{\alpha_i}^2(X) \,  h(x) 
\end{equation}
where $X$ is the messenger scale, $C_3 = 4/3 $ for \SU{3} triplets and $C_2  = 3/4$ for \SU{2} doublets and $C_1 = Y^2 $ represents the hypercharge of the particles.
The function $h(x)$ which is almost flat for most $x$ values is also equal to 1 at $x = 1/2$.  $f$ runs over all the scalars in the theory,
$f = \{Q,~U,D,~L,E,~H_1,~H_2 \}$. Going from three generations to four would not change any of these boundary conditions at the
messenger scale $X$. However, the weak scale spectrum is expected to be completely different from that of the three generation case
due to the presence of large fourth generation Yukawa couplings. It is interesting to study the soft spectrum in the presence of large Yukawa couplings, but a small running scale $\sim \log\left(M_X/M_{\susy{}} \right) \approx 4-5$. To calculate the low energy spectrum we need to compute the RG effects on the soft terms and evaluate the soft mass matrices 
at the weak scale. To that end, we use \indisoft{} which is described in more detail in Appendix \ref{app:indisoft}.

\begin{figure}[h!]
\centering
{\includegraphics[width=0.50\textwidth]{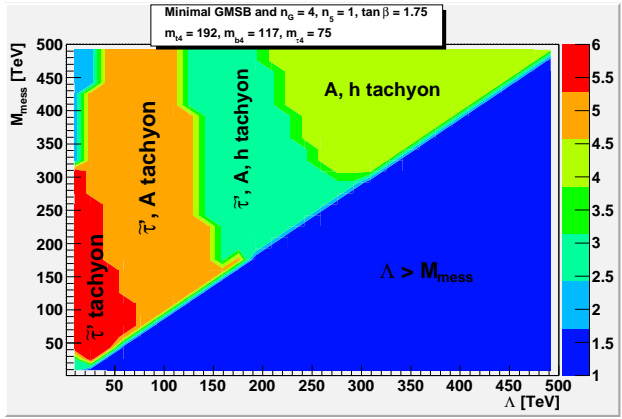}}
\caption{\footnotesize Regions in MMM parameter space $\Lambda$--$M_{\text{mess}}$. The lower-diagonal part is ruled out as $\Lambda > M_{\text{mess}}$. In the upper-diagonal part, from left to right, the first region (red) tachyonic $\tau'$, and the second (orange), third (cyan), fourth (green) do not have consistent radiative electroweak symmetry breaking as indicated by the tachyonic Higgses.}
\label{mmmrun}
\end{figure}

Choosing the masses for the fourth generation to be given by 
\vref{ourlimits} and $\tan\beta=1.75$, we have computed the supersymmetric 
spectrum at the weak scale. We vary $10\text{ TeV} \leq \Lambda, M_{\text{mess}} \leq 500 \text{ TeV}$. The results are characterised by various regions in the parameter space as shown in \vref{mmmrun}. The blue region in the lower diagonal part is ruled out, because the messenger scale $M_\text{mess}$ is smaller than $\Lambda$. The greater part of the parameter space does not have radiative electroweak symmetry breaking as indicated by the tachyonic Higgs scalars (orange, cyan, and green regions) and is thus ruled out. There is, however, a small red region where electroweak symmetry breaking is possible, but the $\widetilde{\tau}'$ is a tachyon.

\medskip

\begin{table}[h!]
\centering
\normalsize
\renewcommand{\arraystretch}{1.4}
\begin{tabular}{|ccrc|ccrc|cr|cr|cr|}
\hline
\multicolumn{4}{|c|}{Higgses [GeV]} & \multicolumn{4}{c}{Gauginos [GeV]} & \multicolumn{6}{|c|}{Squarks \& Sleptons [GeV]}\\
\hline
&$h^0$   & 46.2   &&& $\widetilde{\chi}^0_1$ & 64.3   &&  $\widetilde{u}_L$   & 758.1   &  $\widetilde{t}_1$      & 766.1   &  $\widetilde{t}_1'$      & 722.6\\
&$A^0$   & 507.6  &&& $\widetilde{\chi}^0_2$ & 127.0  &&  $\widetilde{u}_R$   & 735.5   &  $\widetilde{t}_2$      & 639.3   &  $\widetilde{t}_2'$      & 583.8\\ 
&$H^0$   & 532.2  &&& $\widetilde{\chi}^0_3$ & 640.6  &&  $\widetilde{d}_L$   & 761.1   &  $\widetilde{b}_1$      & 725.1   &  $\widetilde{b}_1'$      & 733.4\\ 
&$H^\pm$ & 516.1  &&& $\widetilde{\chi}^0_4$ & 655.1  &&  $\widetilde{d}_R$   & 733.8   &  $\widetilde{b}_2$      & 734.3   &  $\widetilde{b}_2'$      & 525.5\\ 
\cline{1-4}
&        &        &&& $\widetilde{\chi}^\pm_1$ & 126.9  &&  $\widetilde{e}_L$   & 208.3   &  $\widetilde{\tau}_1$   & 208.4   &  $\widetilde{\tau}_1'$   & 320.3\\ 
&        &        &&& $\widetilde{\chi}^\pm_2$ & 652.0  &&  $\widetilde{e}_R$   & 88.1    &  $\widetilde{\tau}_2$   & 87.8    &  $\widetilde{\tau}_2'$   & 193.4\\ 
&        &        &&& $\widetilde{g}$   & 438.4  &&  $\widetilde{\nu}_e$ & 197.2   &  $\widetilde{\nu}_\tau$ & 197.2   &  $\widetilde{\nu}_\tau'$ & 202.7\\ 
\hline
\end{tabular}
\caption{Minimal GMSB spectrum with 4 generations: $n_5=1$ GeV, $M_{\text{mess}}=100$ TeV, $\Lambda=50$ TeV, $\tan\beta=1.75$, \sgn{\mu}=+. $\widetilde{\tau}'$ is tachyonic, $m_{h} = 46.2$ GeV, $m_{\widetilde{G}}=1.2\times10^{-9}$ GeV (gravitino), NLSP is neutralino.}
\label{mmmspectrum}
\end{table}

It is interesting to study in detail the region of the tachyonic
$\widetilde{\tau}'$ , with
\begin{equation*}
10 \text{ TeV} \lesssim \Lambda \lesssim 70 \text{ TeV}, \quad 30 \text{ TeV}
\lesssim M_X \lesssim 300 \text{ TeV}.
\end{equation*} 
A sample spectrum of this region is given in \vref{mmmspectrum} where
we have chosen $\Lambda$ = 50 TeV and $M_{\text{mess}}$ = 100
TeV. Notice that the light Higgs mass does not satisfy the LEP-II constraints for
this point in the parameter space.

\medskip

To trace the reasons for a tachyonic $\widetilde{\tau}'$, let us consider its  mass matrix  which is given by 
\begin{equation}
m^2_{\widetilde{\tau}'}=
\left(
\begin{array}{cc}
m^2_{\sc LL} & m^2_{\sc LR} \\
m^{2\,*}_{\sc LR} & m^2_{\sc RR} 
\end{array} \right)=
{\bf O}^T 
\left(
\begin{array}{cc}
m^2_{\widetilde{\tau}'_1} & 0 \\
0 & m^2_{\widetilde{\tau}'_2}
\end{array} \right) {\bf O},
\end{equation}
where
\begin{eqnarray}
m^2_{\sc LL} &\equiv& m^2_{\widetilde{\tau}'_L} + m_{\tau'}^2 - m^2_Z \cos 2\beta \left(\frac{1}{2} - \sin^2\theta_W \right) \enspace \approx \enspace m^2_{\widetilde{\tau}'_L} \nonumber \\
m^2_{\sc RR} &\equiv& m^2_{\widetilde{\tau}'_R} + m_{\tau'}^2 - m^2_Z \cos 2\beta \sin^2\theta_W \enspace \approx \enspace m^2_{\widetilde{\tau}'_R} \nonumber \\
m^2_{\sc LR} &\equiv& v_d ~(A^*_{\tau'} -\mu ~y_{\tau'} \tan\beta) \enspace \approx \enspace -m_{\tau'} \mu\tan\beta
\label{stauprimemassmatrix}
\end{eqnarray}
where ${\bf O}$ is some mixing matrix which diagonalises the $\widetilde{\tau}'$ mass matrix.  Unless both $\widetilde{\tau}'$ are tachyonic, the condition that there is no tachyon in the spectrum is given by:
\begin{equation}
\label{tachyon}
m^2_{\widetilde{\tau}'_L}   m^2_{\widetilde{\tau}'_R}  - ( m_{\tau'}\,\mu\tan\beta )^2 > 0
\end{equation} 
In the MMM model, it is typical that $\mu$ is  very large, between one  TeV to tens of TeV.  In our analysis, $\mu$ is determined at the weak 
scale by electroweak symmetry breaking conditions. With $m_{\tau'} = 75 $ GeV,  the second term of \vref{tachyon} is quite large.  After
the RG evolution, the soft terms for $\widetilde{\tau}_{L,R}$ take the form \cite{Borzumati:1996qs}:
\begin{eqnarray}
m^2_{\widetilde{\tau}'_L} (qZ) & \approx &  m^2_{\widetilde{L} }(X)  - \tilde{\alpha}_{\tau'} (2 m_{\widetilde{L}}^2 (X) + m_{\widetilde{E}}^2 (X) )   \log(X/qZ) \nonumber \\
m^2_{\widetilde{\tau}'_R} (qZ)   &\approx &  m^2_{\widetilde{E}}(X)  -  2  \tilde{\alpha}_{\tau'}(2 m_{\widetilde{L}}^2 (X) + m_{\widetilde{E}}^2 (X) ) \log(X/qZ)
  \end{eqnarray}
where $\tilde{\alpha}_{\tau'} = y_{\tau'}^2 \big/ (16 \pi^2) \approx  1.1 \times 10^{-3}\times(1+\tan^2\beta)$, \enspace $qZ \approx 1$ TeV is the typical supersymmetric mass scale
close to the weak scale, and $X$ represents the messenger scale as before. Note that the log-factor is extremely small, $\sim~4.6$. Thus, the
effective difference between the high scale and the weak scale sleptons is very small, $\sim (5-10)\%$.  For this reason, it becomes difficult
to maintain \vref{tachyon} positive.  Thus, the  MMM model is perhaps too restrictive for the case of four generations. With modified  boundary conditions at the messenger scale, it might be possible that gauge mediated supersymmetry breaking could lead
to a phenomenologically viable spectrum.  This brings us to the new ideas of general gauge mediation (GGM).   

\subsection{General Gauge Mediation}

In the last couple of years, there has been a paradigm shift in the way we understand gauge mediated supersymmetry
breaking. Ref.~\cite{Meade:2008wd} has introduced the framework of general gauge mediation where the 
starting point is that the hidden sector and the visible MSSM sector should completely decouple in the limit where the gauge couplings are set to zero. The generalisation lies in the set of formulae for the soft terms which span
from models with a weakly coupled messenger sector as well as to the strongly coupled ones. 
The general gauge mediation can allow for different scales for the scalars and the gaugino masses, with $\Lambda_S$
for scalar masses and $\Lambda_G$ for gaugino masses.  In fact, the authors of Ref.~\cite{Abel:2009ve} argue that with this kind of
parametrisation, the GMSB analysis is closer to mSUGRA where the role of $\Lambda_G$ is played by $m_{1/2}$ 
and that of $\Lambda_{S}$ is played by $m_0$.  Such freedom in the choice of scales could lead to a relaxation of the parameter
space. Finally, let us note that this framework of general gauge mediation does not  have any solution to the $\mu$-problem 
as it is not generated through the above interactions.  Typical solutions of the $\mu$-problem could decouple the Higgs
sector from the sleptonic sector which might pave way for the solution to the tachyonic $\widetilde{\tau}$ problem. A more
detailed exploration regarding these issues will be addressed in a forthcoming publication.

\section{Summary and Outlook}
\label{sec:summary}

In the present work, we have revisited in detail the perturbativity
constraints on the fourth generation Yukawa couplings in the minimal
supersymmetric standard model. We have shown that if the fourth
generation masses lie close to their present upper limits (or within
25\% of these limits), it is possible to have perturbativity only up
to $\sim$ 1000 TeV; $\tan\beta$ is confined to be very low in these
models, $\sim 3$.  This makes it hard to reconcile gravity mediated
supersymmetry breaking models with four generations unless one assumes
additional vector-like matter. However, as demonstrated in the
spectrum of the toy model of mSUGRA with four generations, the
presence of a fourth generation can lead to new features like lighter
gluino and squark masses which could be accessible at LHC.

\medskip

With this motivation, we have studied gauge mediated supersymmetry
breaking with four generations. We have explored the already highly
constrained minimal messenger model with the ranges allowed and found
that in most of the parameter space, radiative electroweak symmetry
breaking is not possible. There is a small region where electroweak
symmetry breaking is viable but $\widetilde{\tau}'$ is tachyonic. It
is difficult to accommodate a fourth generation in the highly
constrained minimal messenger model. We speculate that the more
general framework of general gauge mediation would be suited to get
phenomenologically viable mass spectra with four generations.

\medskip

The combination of supersymmetry and four generations seems to lead to
interesting supersymmetric spectroscopy at the weak scale. In the
present work, we have just explored possible supersymmetry breaking
scenarios which can work within this framework. However, four
generational MSSM would have strong implications in various other
sectors like flavour physics, neutrino physics and most interestingly
the Higgs physics in MSSM. These aspects as well as concrete models
within gauge mediation need to be studied. Experimental consequences
of the presence of the fourth generation with SUSY, in B-physics,
D-physics as well as direct detection at LHC needs to be investigated.


\vskip 0.5 cm 
\textbf{Acknowledgements} \\

\noindent
We  acknowledge useful discussions with Ben Allanach,  J\"org Meyer,  Sandip Pakvasa and  Arun Prasath.  SKV acknowledges support by  DST Project  No : SR/S2/HEP-0018/2007
of Govt. of India. RMG acknowledges the support of Department of Science and Technology, India, under the J.C. Bose fellowship  SR/S2/JCB-64/2007.


\clearpage
\newpage

\appendix

\section{The Renormalisation Group Equations}
\label{app:rge}

For completeness, we list the 2-loop renormalisation group equations for the gauge and Yukawa couplings that we have used to generalise \softsusy{} to the case of the MSSM with $n_G$ chiral generations. These RGEs can be derived from the general case \cite{Machacek:1983tz,Machacek:1983fi,Falck:1985aa} and are well-known in the literature \cite{Bjorkman:1985mi,Bagger:1985ig,Cvetic:1985fp,Drees:1987ev,Tanimoto:1987sv,Castano:1993ri,Gunion:1995tp}. Here we are following Ref.~\cite{Castano:1993ri}.

\subsection{The Gauge Couplings}
\label{appendix-gaugecouplings}

The running of the gauge couplings is given by
\begin{equation}
   \frac{dg_i}{dt} = - \frac{1}{16\pi^2} \, b_i^{} g_i^3
    + \frac{g_i^3}{(16\pi^2)^2} \left[ \sum_k b_{ik}^{} g_k^2
   - \trace{ C_{iu}{\bf Y}_u^\dagger{\bf Y}_u^{}
   + C_{id}{\bf Y}_d^\dagger{\bf Y}_d^{}
   + C_{ie}{\bf Y}_e^\dagger{\bf Y}_e^{} } \right],
\end{equation}
where $t=\log(\mu/\mu_0)$, $\mu$ is the renormalisation scale, $\mu_0$ some reference scale, and $i,k=1,2,3$ refer to the gauge groups $\U{1}_Y$, $\SU{2}_L$ and $\SU{3}_c$. At 1-loop, the $b_i$ are given by 
\begin{equation}
b_1 = -\frac{3}{5} - 2n_G, \qquad b_2 = 5 - 2n_G, \qquad b_3 = 9 - 2n_G.
\label{eq:beta-function-coefficients-mssm}
\end{equation}
The fourth generation comes in complete GUT multiplets, and thus most predictions from grand unification are unchanged. In fact, the only modification at 1-loop is that \alphagut{} increases. At 2-loop, the Yukawa couplings enter the RGEs, and the coefficients are given by
\begin{equation}
\left(b \right)_{ik} = 
n_G \left(
\begin{array}{ccc}
\frac{38}{15} & \frac{6}{5} & \frac{88}{15}\\
\frac{2}{5}   & 14          & 8\\
\frac{11}{15} & 3           & \frac{68}{3}\\
\end{array}
\right) 
+ \left(
\begin{array}{ccc}
\frac{9}{25} & \frac{9}{5}  & 0\\
\frac{3}{5}   & \m17        & 0\\
0             & 0           & \m54\\
\end{array}
\right), \qquad
\left(C \right)_{i f} = 
\left(
\begin{array}{ccc}
\frac{26}{5} & \frac{14}{5} & \frac{18}{5}\\
6   & 6          & 2\\
4 & 4           & 0\\
\end{array}
\right)
\end{equation}
where $i,k = 1,2,3$ as before, and $f=u,d,e$.

\subsection{The Yukawa Couplings}
\label{appendix-yukawarge}

The running of the Yukawa couplings is given by
\begin{equation}
\label{yukawarge}
   \frac{d}{dt} {\bf Y}_{u,d,e}^{} = {\bf Y}_{u,d,e}^{} \left(
   \frac{1}{16\pi^2} \bs{\beta}_{u,d,e}^{(1)} + \frac{1}{(16\pi^2)^2}
   \bs{\beta}_{u,d,e}^{(2)} \right).
\end{equation}

At 1-loop, the $\beta$-function coefficients are:
\begin{align}
   \bs{\beta}_u^{(1)} &= 3 {\bf Y}_u^\dagger {\bf Y}_u^{}
                   + {\bf Y}_d^\dagger {\bf Y}_d^{}
                     + 3 \trace{{\bf Y}_u^\dagger{\bf Y}_u^{}}
   - \left( \frac{13}{15}g_1^2 + 3g_2^2 + \frac{16}{3}g_3^2 \right)
\\
   \bs{\beta}_d^{(1)} &= 3 {\bf Y}_d^\dagger {\bf Y}_d^{}
          + {\bf Y}_u^\dagger {\bf Y}_u^{}
          + \trace{3{\bf Y}_d^\dagger{\bf Y}_d^{}
          +{\bf Y}_e^\dagger{\bf Y}_e^{} }
   - \left( \frac{7}{15}g_1^2 + 3g_2^2 + \frac{16}{3}g_3^2 \right)
\\
   \bs{\beta}_e^{(1)} &= 3 {\bf Y}_e^\dagger {\bf Y}_e^{}
       + \trace{ 3{\bf Y}_d^\dagger{\bf Y}_d^{}
          +{\bf Y}_e^\dagger{\bf Y}_e^{} }
   - \left( \frac{9}{5}g_1^2 + 3g_2^2 \right)
\end{align}
Note that the number of generations enters only through the dimensionality of the Yukawa matrices. At 2-loop, $n_G$ is explicitly present, and the $\beta$-function coefficients are:
\begin{eqnarray}
  \bs{\beta}_u^{(2)} &=& - 4({\bf Y}_u^\dagger{\bf Y}_u^{})^2 - 2({\bf Y}_d^\dagger{\bf Y}_d^{})^2 - 2{\bf Y}_d^\dagger {\bf Y}_d^{}{\bf Y}_u^\dagger{\bf Y}_u^{} - 9\trace{{\bf Y}_u^\dagger{\bf Y}_u^{}} {\bf Y}_u^\dagger{\bf Y}_u^{} \nonumber \\
  &&\mbox{}- \trace{3{\bf Y}_d^\dagger{\bf Y}_d^{}+{\bf Y}_e^\dagger{\bf Y}_e^{}}{\bf Y}_d^\dagger{\bf Y}_d^{} - 3\trace{3({\bf Y}_u^\dagger{\bf Y}_u^{})^2+{\bf Y}_d^\dagger{\bf Y}_d^{}{\bf Y}_u^\dagger{\bf Y}_u^{}} \nonumber \\
  &&\mbox{}+ (\frac{2}{5}g_1^2 + 6g_2^2){\bf Y}_u^\dagger{\bf Y}_u^{} + (\frac{2}{5}g_1^2){\bf Y}_d^\dagger{\bf Y}_d^{} + (\frac{4}{5}g_1^2+16g_3^2)\trace{{\bf Y}_u^\dagger{\bf Y}_u^{}} \nonumber \\
  &&\mbox{}+ (\frac{26}{15}n_G+\frac{403}{450})g_1^4 + (6n_G-\frac{21}{2})g_2^4 + (\frac{32}{3}n_G-\frac{304}{9})g_3^4 \nonumber \\
  &&\mbox{}+ g_1^2g_2^2 + \frac{136}{15}g_1^2g_3^2 + 8g_2^2g_3^2 \\
  \bs{\beta}_d^{(2)} &=& - 4({\bf Y}_d^\dagger{\bf Y}_d^{})^2 - 2({\bf Y}_u^\dagger{\bf Y}_u^{})^2 - 2{\bf Y}_u^\dagger {\bf Y}_u^{}{\bf Y}_d^\dagger{\bf Y}_d^{} - 3\trace{{\bf Y}_u^\dagger{\bf Y}_u^{}} {\bf Y}_u^\dagger{\bf Y}_u^{} \nonumber \\
  &&\mbox{}- 3\trace{3{\bf Y}_d^\dagger{\bf Y}_d^{}+{\bf Y}_e^\dagger {\bf Y}_e^{}}{\bf Y}_d^\dagger{\bf Y}_d^{} - 3\trace{3({\bf Y}_d^\dagger{\bf Y}_d^{})^2+({\bf Y}_e^\dagger {\bf Y}_e^{})^2+{\bf Y}_d^\dagger{\bf Y}_d^{}{\bf Y}_u^\dagger {\bf Y}_u^{}} \nonumber \\
  &&\mbox{}+ (\frac{4}{5}g_1^2){\bf Y}_u^\dagger{\bf Y}_u^{} + (\frac{4}{5}g_1^2+6g_2^2){\bf Y}_d^\dagger{\bf Y}_d^{} + (-\frac{2}{5}g_1^2+16g_3^2)\trace{{\bf Y}_d^\dagger {\bf Y}_d^{}} + (\frac{6}{5}g_1^2)\trace{{\bf Y}_e^\dagger{\bf Y}_e^{}} \nonumber \\
  &&\mbox{}+ (\frac{14}{15}n_G+\frac{7}{18})g_1^4 + (6n_G-\frac{21}{2})g_2^4 + (\frac{32}{3}n_G-\frac{304}{9})g_3^4 + g_1^2g_2^2  + \frac{8}{9} g_1^2g_3^2 + 8g_2^2g_3^2 \\
  \bs{\beta}_e^{(2)} &=& - 4({\bf Y}_e^\dagger{\bf Y}_e^{})^2 - 3\trace{3{\bf Y}_d^\dagger{\bf Y}_d^{} + {\bf Y}_e^\dagger{\bf Y}_e^{}}{\bf Y}_e^\dagger{\bf Y}_e^{} - 3{\rm Tr}\left\{3({\bf Y}_d^\dagger{\bf Y}_d^{})^2+({\bf Y}_e^\dagger {\bf Y}_e^{})^2 \right.\nonumber \\
  &&\mbox{}\left. +{\bf Y}_d^\dagger{\bf Y}_d^{}{\bf Y}_u^\dagger{\bf Y}_u^{}\right\} + (6g_2^2){\bf Y}_e^\dagger{\bf Y}_e^{} + (\frac{6}{5}g_1^2)\trace{{\bf Y}_e^\dagger{\bf Y}_e^{}} + (-\frac{2}{5}g_1^2+16g_3^2)\trace{{\bf Y}_d^\dagger{\bf Y}_d^{}} \nonumber \\
  &&\mbox{}+ (\frac{18}{5}n_G+\frac{27}{10})g_1^4 + (6n_G-\frac{21}{2})g_2^4 + \frac{9}{5}g_1^2g_2^2
\end{eqnarray}


\subsection{Approximate Upper Bounds}
\label{appendix-upperbounds}

An approximate upper bound on the fourth generation $t'$ can be
obtained in the limit where $m_{t'}$ dominates over all the other
masses, $m_{t'} \gg \{ m_{t},~m_{b'},~m_{\tau'} \}$ purely from
perturbative requirements. To see this, let us rewrite
\vref{yukawarge} in terms of the dominant Yukawa couplings:
\begin{eqnarray}
\label{rge}
\beta_{h_{t'}} &=& h_{t'} \left( 6 h_{t'}^2 + h_{b'}^2 + 3 h_t^2 - {16 \over 3} g_3^2 - 3 g_2^2 - {13 \over 15} g_1^2 \right) \nonumber \\
\beta_{h_{b'}} &=& h_{b'} \left(6 h_{b'}^2 + h_{t'}^2 + h_{\tau'}^2 - {16 \over 3} g_3^2 - 3 g_2^2 - { 7 \over 15} g_1^2  \right) \nonumber \\
\beta_{h_{\tau'}} &=& h_{\tau'} \left(3 h_{b'}^2 + 4  h_{\tau'}^2  - 3 g_2^2 - { 9 \over 5} g_1^2  \right) \nonumber \\
\beta_{h_t} &=& h_t \left( 3 h_{t'}^2 + 6 h_t^2 - {16 \over 3} g_3^2 - 3 g_2^2 - {13 \over 15} g_1^2 \right)
\end{eqnarray}
These equations are non-linear and coupled and obviously, should be
solved numerically. However an analytical estimate can be obtained in
the above mentioned limit $m_{t'} \gg \{ m_{t},~m_{b'},~m_{\tau'} \}$.
In this limit the solution for $h_{t'}$ can be written as
\cite{Ibanez:1983di} :
\begin{equation}
\label{yukawasolution}
k_{t'} (z) = {E (z) k_{t'} (0) \over ( 1 + 6 k_{t'}(0) F (z)) },
\end{equation}
where $z = 2 \log (M_X/M_Z)$ with $M_X$ denoting the high scale and
$M_Z$, the weak scale. $k_{t'} = h_{t'}^2 / (4 \pi)$ and
\begin{eqnarray}
E(z) & =& \left(1 + \tilde{b}_1 {\alpha_{1}(0) \over 4 \pi }z\right)^{{13\over
9 \tilde{b}_1}} \left(1 + \tilde{b}_2 {\alpha_{2}(0) \over 4 \pi }z\right)^{{3\over
\tilde{b}_2}} \left(1 + \tilde{b}_3 {\alpha_{3}(0) \over 4 \pi }z\right)^{{16 \over 3
\tilde{b}_3}} \nonumber \\ F(z) &=& \int_{0}^{z} E(z') dz'
\label{eq:solyukawa}
\end{eqnarray}
$k_{t'}(0)$ is the value of the $t'$ Yukawa (squared) at the high
scale, $M_X$. These are the same expressions one obtains for the top
quark Yukawa coupling within three generations. In four generations,
the only difference is the $\beta$-functions of the gauge couplings.
 In \vref{eq:solyukawa}, $\tilde{b}_i = - b_i$, and 
from \vref{eq:beta-function-coefficients-mssm}, $\tilde{b}_3 = -1$,
$\tilde{b}_2 = 3$, $\tilde{b}_1 = 43/3 $. If the Yukawa coupling becomes very large,
 from \vref{yukawasolution}, we can derive an upper bound
on the (square) of the Yukawa of the top-prime in the limit $k_{t'}(0) \to \infty$:
\begin{equation}
k_{t'} (z) \sim {E (z) \over 6~ F (z) }
\label{upperbound}
\end{equation}
In \vref{topprimelimit} we present the upper limits on the $m_{t'}$ in
this approximate limit using \vref{upperbound}. We see that the
present limit on the $m_{t'}$ readily rules out perturbativity up to
the Planck scale or even the GUT scale.  If we demand
perturbativity beyond 1-100 TeV, $m_{t'}$ is forced to be close to its
experimental present lower bound or below. Beyond $10^6$ GeV the
present limit already rules out perturbative a Yukawa coupling for the
$t'$.  As we have seen in the main text, demanding perturbativity of
the $b'$ and $\tau'$ Yukawa couplings would put further stringent
constraints on the scale $M_X$.

\begin{table}[htdp]
\begin{center}
\begin{tabular}{|c|c|c|c|c|c|c|c|}
\hline
$M_X$ (GeV) & $10^{18}$ & $10^{16}$ & $10^{11}$ & $10^{6}$ & $10^{3} $
\\
\hline 
$ m_{t'}$ (GeV) &211.3&212.8&222.2&264.2&437.7 \\
\hline
\end{tabular}
\end{center}
\caption{The approximate upper limits on $m_{t'}$ in the limit where $m_{t'}$ 
dominates over all the masses, for various high scales, $M_X$.}
\label{topprimelimit}
\end{table}


\section{\indisoft{}}
\label{app:indisoft}

In this appendix, we summarise the most important features of the software tool that we have developed to calculate soft spectra in the MSSM with four generations. More details will be presented in a separate publication.

\medskip

\indisoft{} is based on \softsusy{} 3.0.9.~\cite{Allanach:2001kg}. We have preserved the original \cpp{} \textit{class} structure and extended the functionality of the individual \textit{classes} that handle the calculations. It is beyond the scope of the present appendix to describe \softsusy{} in detail, and we refer for that to its manual \cite{Allanach:2001kg}. 

\medskip

The fermion masses and gauge coupling constants are stored in the \textit{class} \code{QedQcd} which we have extended to comprise also the fourth generation fermion masses. The light fermion masses and the couplings that are entered at different scales (depending on where they are experimentally known) are then run to $M_Z$ using the \textit{class} \code{RGE} that implements the renormalisation group running for every class that \textit{derives} from it, in this case for \code{QedQcd}. For the heavy fermions, we have generalised the functions that calculate the pole mass from the running one and vice versa. From now on we will assume that we are working in a framework (like mSUGRA or gauge mediation) where the soft masses are obtained from boundary conditions set at a higher scale. The \textit{class} \code{MssmSusy} is derived from \code{RGE} and contains the supersymmetric parameters of the theory; the Yukawa matrices have been promoted to $4\times4$ matrices to include the fourth generation. The $\beta$-functions of \code{MssmSusy} have been generalised to the case of four generations (see Appendix \ref{app:rge}). The \textit{class} \code{SoftParsMssm} is derived from \code{MssmSusy} and contains the soft supersymmetry-breaking terms of the MSSM; the soft mass matrices and the trilinear couplings have been generalised; the supersymmetric parameters and their RGE evolution are inherited from \code{MssmSusy}; the $\beta$-functions of \code{SoftParsMssm} for the soft terms have been generalised. The \textit{class} \code{MssmSoftsusy} derives from \code{SoftParsMssm} and organises the actual calculation of the spectrum. \code{QedQcd} is used to initialise/guess the \susy{} parameters at the scale $m_t$ which are then run by \code{RGE} to the high scale $M_X$ where the boundary conditions on the soft terms are imposed. Note that for the gauge and Yukawa couplings, we use 2-loop RGEs (see Appendix \ref{app:rge}), and 1-loop RGEs for the rest. \code{MssmSoftsusy} is then run back to $M_\susy{}$ ($=M_Z$ for the first iteration), where the electroweak symmetry breaking conditions are checked and the physical spectrum is calculated (at tree-level for the first iteration, and later at 1-loop). The \textit{class} \code{sPhysical} that calculates the physical masses and stores the results has been generalised to handle four generations. The \textit{class} \code{drBarPars} that derives from \code{sPhysical} and manages the $\overline{\text{DR}}$ parameters for the calculation has been generalised. The tadpoles, 1-loop radiative corrections to the Higgses, squarks, sleptons, and the threshold corrections to the gauge couplings have been generalised to include contributions from the fourth generation fermions. The aforementioned steps are iterated until satisfactory convergence is achieved. 

\medskip

As a result of our work, we found some minor typos and bugs\footnote{\softsusy{} 3.0 to 3.0.9 did not correctly indicate the regions where no electroweak symmetry breaking is possible, and in release 3.0.9, there were some typos in the formulae for the radiative corrections that affected the calculation of the physical masses at the subpercentage level.} in \softsusy{} that have been fixed in subsequent versions. We have rewritten the linear algebra classes from scratch\footnote{During the final stages of this publication, we became aware of the \softsusy{} 3.1 release in which the linear algebra classes have been rewritten by D.~Grellscheid. We have not compared our changes to his.}, replacing the pointer constructions used to represent vectors/matrices and the algorithms operating on them by the Standard Template Library (\texttt{STL}) containers and algorithms. In addition to the presently available mSUGRA and minimal GMSB boundary conditions, we have implemented a model of general gauge mediation along the lines of Ref.~\cite{Abel:2009ve}. We have linked \Root{} \cite{Brun:1997pa} to our programs to generate plots both interactively and in batch mode. In future, we plan to extend \indisoft{} by right-handed neutrinos. Due to space limitations, we must refrain from discussing all the changes we have made.


\clearpage
\newpage

\bibliography{mybibliography}

\providecommand{\href}[2]{#2}\begingroup\raggedright\begin{thebibliography}{10}

\bibitem{Abazov:2009ii}
{\bf D\O} Collaboration, V.~M. Abazov {\em et al.}, ``{Observation of Single
  Top-Quark Production},'' {\em Phys. Rev. Lett.} {\bf 103} (2009) 092001,
\href{http://www.arXiv.org/abs/0903.0850}{{\tt 0903.0850}}.

\bibitem{Aaltonen:2009jj}
{\bf CDF} Collaboration, T.~Aaltonen {\em et al.}, ``{First Observation of
  Electroweak Single Top Quark Production},'' {\em Phys. Rev. Lett.} {\bf 103}
  (2009) 092002,
\href{http://www.arXiv.org/abs/0903.0885}{{\tt 0903.0885}}.

\bibitem{Alwall:2006bx}
J.~Alwall {\em et al.}, ``{Is V(tb) = 1?},'' {\em Eur. Phys. J.} {\bf C49}
  (2007) 791--801,
\href{http://www.arXiv.org/abs/hep-ph/0607115}{{\tt hep-ph/0607115}}.

\bibitem{Kribs:2007nz}
G.~D. Kribs, T.~Plehn, M.~Spannowsky, and T.~M.~P. Tait, ``{Four generations
  and Higgs physics},'' {\em Phys. Rev.} {\bf D76} (2007) 075016,
\href{http://www.arXiv.org/abs/0706.3718}{{\tt 0706.3718}}.

\bibitem{Bobrowski:2009ng}
M.~Bobrowski, A.~Lenz, J.~Riedl, and J.~Rohrwild, ``{How much space is left for
  a new family of fermions?},'' {\em Phys. Rev.} {\bf D79} (2009) 113006,
\href{http://www.arXiv.org/abs/0902.4883}{{\tt 0902.4883}}.

\bibitem{Chanowitz:2009mz}
M.~S. Chanowitz, ``{Bounding CKM Mixing with a Fourth Family},'' {\em Phys.
  Rev.} {\bf D79} (2009) 113008,
\href{http://www.arXiv.org/abs/0904.3570}{{\tt 0904.3570}}.

\bibitem{Holdom:2009rf}
B.~Holdom {\em et al.}, ``{Four Statements about the Fourth Generation},''
\href{http://www.arXiv.org/abs/0904.4698}{{\tt 0904.4698}}.

\bibitem{Novikov:2009kc}
V.~A. Novikov, A.~N. Rozanov, and M.~I. Vysotsky, ``{Once more on extra
  quark-lepton generations and precision measurements},''
\href{http://www.arXiv.org/abs/0904.4570}{{\tt 0904.4570}}.

\bibitem{Hung:2007ak}
P.~Q. Hung and M.~Sher, ``{Experimental constraints on fourth generation quark
  masses},'' {\em Phys. Rev.} {\bf D77} (2008) 037302,
\href{http://www.arXiv.org/abs/0711.4353}{{\tt 0711.4353}}.

\bibitem{Frampton:1999xi}
P.~H. Frampton, P.~Q. Hung, and M.~Sher, ``{Quarks and leptons beyond the third
  generation},'' {\em Phys. Rept.} {\bf 330} (2000) 263,
\href{http://www.arXiv.org/abs/hep-ph/9903387}{{\tt hep-ph/9903387}}.

\bibitem{Nielsen:1995gx}
H.~B. Nielsen, A.~V. Novikov, V.~A. Novikov, and M.~I. Vysotsky, ``{Higgs
  potential bounds on extra quark - lepton generations},'' {\em Phys. Lett.}
  {\bf B374} (1996) 127--130,
\href{http://www.arXiv.org/abs/hep-ph/9511340}{{\tt hep-ph/9511340}}.

\bibitem{Pirogov:1998tj}
Y.~F. Pirogov and O.~V. Zenin, ``{Two-loop renormalization group restrictions
  on the standard model and the fourth chiral family},'' {\em Eur. Phys. J.}
  {\bf C10} (1999) 629--638,
\href{http://www.arXiv.org/abs/hep-ph/9808396}{{\tt hep-ph/9808396}}.

\bibitem{Goldberg:1985gj}
H.~Goldberg, ``{The fourth generation and N=1 Supergravity},'' {\em Phys.
  Lett.} {\bf B165} (1985)
292.

\bibitem{Enqvist:1985ct}
K.~Enqvist, D.~V. Nanopoulos, and F.~Zwirner, ``{The Fourth Generation in
  Supergravity},'' {\em Phys. Lett.} {\bf B164} (1985)
321.

\bibitem{Arnowitt:1987xk}
R.~L. Arnowitt and P.~Nath, ``{Fourth Generation and Nucleon Decay in
  Supersymmetric Theories},'' {\em Phys. Rev.} {\bf D36} (1987)
3423--3428.

\bibitem{Drees:1987ev}
M.~Drees, K.~Enqvist, and D.~V. Nanopoulos, ``{No Future for the Fourth
  Generation?},'' {\em Nucl. Phys.} {\bf B294} (1987)
1.

\bibitem{Gunion:1994zm}
J.~F. Gunion, D.~W. McKay, and H.~Pois, ``{Gauge coupling unification and the
  minimal SUSY model: a Fourth generation below the top?},'' {\em Phys. Lett.}
  {\bf B334} (1994) 339--347,
\href{http://www.arXiv.org/abs/hep-ph/9406249}{{\tt hep-ph/9406249}}.

\bibitem{Gunion:1995tp}
J.~F. Gunion, D.~W. McKay, and H.~Pois, ``{A Minimal four family supergravity
  model},'' {\em Phys. Rev.} {\bf D53} (1996) 1616--1647,
\href{http://www.arXiv.org/abs/hep-ph/9507323}{{\tt hep-ph/9507323}}.

\bibitem{Dubicki:2003am}
J.~E. Dubicki and C.~D. Froggatt, ``{Supersymmetric grand unification with a
  fourth generation?},'' {\em Phys. Lett.} {\bf B567} (2003) 46--52,
\href{http://www.arXiv.org/abs/hep-ph/0305007}{{\tt hep-ph/0305007}}.

\bibitem{Murdock:2008rx}
Z.~Murdock, S.~Nandi, and Z.~Tavartkiladze, ``{Perturbativity and a Fourth
  Generation in the MSSM},'' {\em Phys. Lett.} {\bf B668} (2008) 303--307,
\href{http://www.arXiv.org/abs/0806.2064}{{\tt 0806.2064}}.

\bibitem{Hung:1997zj}
P.~Q. Hung, ``{Minimal SU(5) resuscitated by long-lived quarks and leptons},''
  {\em Phys. Rev. Lett.} {\bf 80} (1998) 3000--3003,
\href{http://www.arXiv.org/abs/hep-ph/9712338}{{\tt hep-ph/9712338}}.

\bibitem{Hou:2008xd}
W.-S. Hou, ``{CP Violation and Baryogenesis from New Heavy Quarks},'' {\em
  Chin. J. Phys.} {\bf 47} (2009) 134,
\href{http://www.arXiv.org/abs/0803.1234}{{\tt 0803.1234}}.

\bibitem{Fok:2008yg}
R.~Fok and G.~D. Kribs, ``{Four Generations, the Electroweak Phase Transition,
  and Supersymmetry},'' {\em Phys. Rev.} {\bf D78} (2008) 075023,
\href{http://www.arXiv.org/abs/0803.4207}{{\tt 0803.4207}}.

\bibitem{Kikukawa:2009mu}
Y.~Kikukawa, M.~Kohda, and J.~Yasuda, ``{The strongly coupled fourth family and
  a first-order electroweak phase transition (I) quark sector},''
\href{http://www.arXiv.org/abs/0901.1962}{{\tt 0901.1962}}.

\bibitem{DePree:2009ed}
E.~De~Pree, G.~Marshall, and M.~Sher, ``{The Fourth Generation t-prime in
  Extensions of the Standard Model},'' {\em Phys. Rev.} {\bf D80} (2009)
  037301,
\href{http://www.arXiv.org/abs/0906.4500}{{\tt 0906.4500}}.

\bibitem{Burdman:2007sx}
G.~Burdman and L.~Da~Rold, ``{Electroweak Symmetry Breaking from a Holographic
  Fourth Generation},'' {\em JHEP} {\bf 12} (2007) 086,
\href{http://www.arXiv.org/abs/0710.0623}{{\tt 0710.0623}}.

\bibitem{BorstnikBracic:2006xc}
A.~Borstnik~Bracic, M.~Breskvar, D.~Lukman, and N.~S. Mankoc~Borstnik, ``{A new
  understanding of fermion masses from the unified theory of spins and
  charges},''
\href{http://www.arXiv.org/abs/hep-ph/0606224}{{\tt hep-ph/0606224}}.

\bibitem{Stremnitzer:1987zp}
H.~Stremnitzer and J.~C. Pati, ``{New Crucial Tests of Compositeness of the
  Third and a Possible Fourth Family in e+ e- Colliders},'' {\em Phys. Lett.}
  {\bf B196} (1987)
240.

\bibitem{Frandsen:2009fs}
M.~T. Frandsen, I.~Masina, and F.~Sannino, ``{Fourth Lepton Family is Natural
  in Technicolor},''
\href{http://www.arXiv.org/abs/0905.1331}{{\tt 0905.1331}}.

\bibitem{Antipin:2009ks}
O.~Antipin, M.~Heikinheimo, and K.~Tuominen, ``{Natural fourth generation of
  leptons},''
\href{http://www.arXiv.org/abs/0905.0622}{{\tt 0905.0622}}.

\bibitem{Holdom:1986rn}
B.~Holdom, ``{Heavy Quarks and Electroweak Symmetry Breaking},'' {\em Phys.
  Rev. Lett.} {\bf 57} (1986)
2496.

\bibitem{Hill:1990ge}
C.~T. Hill, M.~A. Luty, and E.~A. Paschos, ``{Electroweak symmetry breaking by
  fourth generation condensates and the neutrino spectrum},'' {\em Phys. Rev.}
  {\bf D43} (1991)
3011--3025.

\bibitem{King:1990he}
S.~F. King, ``{Is Electroweak Symmetry Broken by a Fourth Family of Quarks?},''
  {\em Phys. Lett.} {\bf B234} (1990)
108--112.

\bibitem{Kramer:1981sq}
G.~Kramer and I.~Montvay, ``{Radiative Quark Mass Generation and a Fourth Quark
  Family},'' {\em Zeit. Phys.} {\bf C11} (1981)
159.

\bibitem{Kagan:1989fp}
A.~L. Kagan, ``{Radiative Quark Mass and Mixing Hierarchies from Supersymmetric
  Models with a Fourth Mirror Family},'' {\em Phys. Rev.} {\bf D40} (1989)
173.

\bibitem{Sher:1992yr}
M.~Sher and Y.~Yuan, ``{Cosmological bounds on the lifetime of a fourth
  generation charged lepton},'' {\em Phys. Lett.} {\bf B285} (1992)
336--342.

\bibitem{Fritzsch:1992bv}
H.~Fritzsch, ``{Light neutrinos, nonuniversality of the leptonic weak
  interaction and a fourth massive generation},'' {\em Phys. Lett.} {\bf B289}
  (1992)
92--96.

\bibitem{Hill:1989vn}
C.~T. Hill and E.~A. Paschos, ``{A Naturally Heavy Fourth Generation
  Neutrino},'' {\em Phys. Lett.} {\bf B241} (1990)
96.

\bibitem{Babu:2009aq}
K.~S. Babu, S.~Nandi, and Z.~Tavartkiladze, ``{New Mechanism for Neutrino Mass
  Generation and Triply Charged Higgs Bosons at the LHC},'' {\em Phys. Rev.}
  {\bf D80} (2009) 071702,
\href{http://www.arXiv.org/abs/0905.2710}{{\tt 0905.2710}}.

\bibitem{Drees:1986ug}
M.~Drees, K.~Enqvist, and D.~V. Nanopoulos, ``{The Fourth Generation in
  Superstring Models},'' {\em Phys. Lett.} {\bf B189} (1987)
321.

\bibitem{Amsler:2008zzb}
{\bf Particle Data Group} Collaboration, C.~Amsler {\em et al.}, ``{Review of
  particle physics},'' {\em Phys. Lett.} {\bf B667} (2008)
1.

\bibitem{:2008nf}
{\bf CDF} Collaboration, T.~Aaltonen {\em et al.}, ``{Search for Heavy Top-like
  Quarks Using Lepton Plus Jets Events in 1.96-TeV $p \bar{p}$ Collisions},''
  {\em Phys. Rev. Lett.} {\bf 100} (2008) 161803,
\href{http://www.arXiv.org/abs/0801.3877}{{\tt 0801.3877}}.

\bibitem{Abachi:1995ms}
{\bf D\O} Collaboration, S.~Abachi {\em et al.}, ``{Top quark search with the
  D\O\ 1992 - 1993 data sample},'' {\em Phys. Rev.} {\bf D52} (1995)
4877--4919.

\bibitem{Aaltonen:2007je}
{\bf CDF} Collaboration, T.~Aaltonen {\em et al.}, ``{Search for New Particles
  Leading to $Z +$ jets Final States in $p \bar{p}$ Collisions at $\sqrt{s}$ =
  1.96- TeV},'' {\em Phys. Rev.} {\bf D76} (2007) 072006,
\href{http://www.arXiv.org/abs/0706.3264}{{\tt 0706.3264}}.

\bibitem{Acosta:2002ju}
{\bf CDF} Collaboration, D.~E. Acosta {\em et al.}, ``{Search for long-lived
  charged massive particles in $\bar{p}p$ collisions at $\sqrt{s} = 1.8$
  TeV},'' {\em Phys. Rev. Lett.} {\bf 90} (2003) 131801,
\href{http://www.arXiv.org/abs/hep-ex/0211064}{{\tt hep-ex/0211064}}.

\bibitem{Achard:2001qw}
{\bf L3} Collaboration, P.~Achard {\em et al.}, ``{Search for heavy neutral and
  charged leptons in $e^{+} e^{-}$ annihilation at LEP},'' {\em Phys. Lett.}
  {\bf B517} (2001) 75--85,
\href{http://www.arXiv.org/abs/hep-ex/0107015}{{\tt hep-ex/0107015}}.

\bibitem{Abreu:1991pr}
{\bf DELPHI} Collaboration, P.~Abreu {\em et al.}, ``{Searches for heavy
  neutrinos from Z decays},'' {\em Phys. Lett.} {\bf B274} (1992)
230--238.

\bibitem{Rattazzi:1995gk}
R.~Rattazzi and U.~Sarid, ``{The Unified minimal supersymmetric model with
  large Yukawa couplings},'' {\em Phys. Rev.} {\bf D53} (1996) 1553--1585,
\href{http://www.arXiv.org/abs/hep-ph/9505428}{{\tt hep-ph/9505428}}.

\bibitem{Ananthanarayan:1991xp}
B.~Ananthanarayan, G.~Lazarides, and Q.~Shafi, ``{Top mass prediction from
  supersymmetric guts},'' {\em Phys. Rev.} {\bf D44} (1991)
1613--1615.

\bibitem{Allanach:2001kg}
B.~C. Allanach, ``{SOFTSUSY: A C++ program for calculating supersymmetric
  spectra},'' {\em Comput. Phys. Commun.} {\bf 143} (2002) 305--331,
\href{http://www.arXiv.org/abs/hep-ph/0104145}{{\tt hep-ph/0104145}}.

\bibitem{Litsey:2009rp}
S.~Litsey and M.~Sher, ``{Higgs Masses in the Four Generation MSSM},'' {\em
  Phys. Rev.} {\bf D80} (2009) 057701,
\href{http://www.arXiv.org/abs/0908.0502}{{\tt 0908.0502}}.

\bibitem{Giudice:1998bp}
G.~F. Giudice and R.~Rattazzi, ``{Theories with gauge-mediated supersymmetry
  breaking},'' {\em Phys. Rept.} {\bf 322} (1999) 419--499,
\href{http://www.arXiv.org/abs/hep-ph/9801271}{{\tt hep-ph/9801271}}.

\bibitem{Drees:2004jm}
M.~Drees, R.~Godbole, and P.~Roy, ``{Theory and phenomenology of sparticles: An
  account of four-dimensional N=1 supersymmetry in high energy physics},''.
  Hackensack, USA: World Scientific (2004) 555 p.

\bibitem{Borzumati:1996qs}
F.~Borzumati, ``{On the minimal messenger model},''
\href{http://www.arXiv.org/abs/hep-ph/9702307}{{\tt hep-ph/9702307}}.

\bibitem{Meade:2008wd}
P.~Meade, N.~Seiberg, and D.~Shih, ``{General Gauge Mediation},'' {\em Prog.
  Theor. Phys. Suppl.} {\bf 177} (2009) 143--158,
\href{http://www.arXiv.org/abs/0801.3278}{{\tt 0801.3278}}.

\bibitem{Abel:2009ve}
S.~Abel, M.~J. Dolan, J.~Jaeckel, and V.~V. Khoze, ``{Phenomenology of Pure
  General Gauge Mediation},''
\href{http://www.arXiv.org/abs/0910.2674}{{\tt 0910.2674}}.

\bibitem{Machacek:1983tz}
M.~E. Machacek and M.~T. Vaughn, ``{Two Loop Renormalization Group Equations in
  a General Quantum Field Theory. 1. Wave Function Renormalization},'' {\em
  Nucl. Phys.} {\bf B222} (1983)
83.

\bibitem{Machacek:1983fi}
M.~E. Machacek and M.~T. Vaughn, ``{Two Loop Renormalization Group Equations in
  a General Quantum Field Theory. 2. Yukawa Couplings},'' {\em Nucl. Phys.}
  {\bf B236} (1984)
221.

\bibitem{Falck:1985aa}
N.~K. Falck, ``{Renormalization Group Equations for Softly Broken
  Supersymmetry: The Most General Case},'' {\em Z. Phys.} {\bf C30} (1986)
247.

\bibitem{Bjorkman:1985mi}
J.~E. Bjorkman and D.~R.~T. Jones, ``{The Unification Mass, $\sin^2\theta_w$
  and $m_b/m_\tau$ in Nonminimal Supersymmetric SU(5)},'' {\em Nucl. Phys.}
  {\bf B259} (1985)
533.

\bibitem{Bagger:1985ig}
J.~Bagger, S.~Dimopoulos, and E.~Masso, ``{Renormalization Group Constraints in
  Supersymmetric Theories},'' {\em Phys. Rev. Lett.} {\bf 55} (1985)
920.

\bibitem{Cvetic:1985fp}
M.~Cvetic and C.~R. Preitschopf, ``{Heavy Families and N=1 Supergravity Within
  the Standard Model},'' {\em Nucl. Phys.} {\bf B272} (1986)
490.

\bibitem{Tanimoto:1987sv}
M.~Tanimoto, Y.~Suetake, and K.~Senba, ``{Fritzsch Mass Matrix with the Fourth
  Generation and the Renormalization Group Equations},'' {\em Phys. Rev.} {\bf
  D36} (1987)
2119.

\bibitem{Castano:1993ri}
D.~J. Castano, E.~J. Piard, and P.~Ramond, ``{Renormalization group study of
  the Standard Model and its extensions. 2. The Minimal supersymmetric Standard
  Model},'' {\em Phys. Rev.} {\bf D49} (1994) 4882--4901,
\href{http://www.arXiv.org/abs/hep-ph/9308335}{{\tt hep-ph/9308335}}.

\bibitem{Ibanez:1983di}
L.~E. Ibanez and C.~Lopez, ``{N=1 Supergravity, the Weak Scale and the
  Low-Energy Particle Spectrum},'' {\em Nucl. Phys.} {\bf B233} (1984)
511.

\bibitem{Brun:1997pa}
R.~Brun and F.~Rademakers, ``{ROOT: An object oriented data analysis
  framework},'' {\em Nucl. Instrum. Meth.} {\bf A389} (1997)
81--86.

\end{thebibliography}\endgroup

\bibliographystyle{./utphys}

\end{document}